\documentclass[aps,prl,twocolumn,superscriptaddress, preprintnumbers,showkeys,nofootinbib,notitlepage,longbibliography]{revtex4-1}

\usepackage{graphicx}

\usepackage[english]{babel}
\usepackage{xcolor}
\usepackage{mathtools}
\usepackage{pdfsync}
\usepackage{amssymb}
\usepackage{amsmath}

\usepackage[normalem]{ulem}

\usepackage{mathrsfs}
\usepackage{hyperref}
\usepackage{slashed}

\def \matrix #1 {\left(\begin{array}{cc} #1 \end{array}\right)}

\def\II{\hbox{{1}\kern-.25em\hbox{l}}}

\newcommand \widebar [1] {\overline{#1}}

\newcommand \ket [1] {|{#1}\rangle}
\newcommand \bra [1] {\langle {#1}|}

\newcommand{\f}[1]{\mathcal{#1}}

\newcommand{\MS}{$\widebar{\mathrm{MS}}\,\,$}

\begin{document}

\title{Deeply-virtual Compton scattering at the next-to-next-to-leading order}

\author{V. M. Braun}
\email{vladimir.braun@ur.de}
\affiliation{Institut f\"ur Theoretische Physik, Universit\"at Regensburg, D-93040 Regensburg, Germany}

\author{Yao Ji}
\email{ yao.ji@tum.de}
\affiliation{Physik Department T31, James-Franck-Stra\ss e 1, Technische Universit\"at M\"unchen, D-85748 Garching, Germany}

\author{Jakob Schoenleber}
\email{Jakob.Schoenleber@physik.uni-regensburg.de}
\affiliation{Institut f\"ur Theoretische Physik, Universit\"at Regensburg, D-93040 Regensburg, Germany}

\date{\today}

\begin{abstract}
Deeply-virtual Compton scattering gives access to the generalized parton distributions that encode the information on the 
transverse position of quarks and gluons in the proton with dependence on their longitudinal momentum.
In anticipation of the high-precision experimental data in a broad kinematic range from the Electron-Ion Collider,
we have calculated the two-loop, next-to-next-to-leading order (NNLO) DVCS coefficient functions associated with the dominant 
Compton form factors $\f H$ and $\f E$ at large energies. The NNLO correction to the imaginary part of $\f H$ appears to be rather large, 
up to factor two at the input scale $Q^2=4$~GeV$^2$ for simple GPD models, due to a cancellation between quark and gluon contributions.    
\end{abstract}

\preprint{TUM-HEP-1407/22}

\maketitle

%
\section{Introduction}
%
The physics program of the planned  Electron-Ion Collider (EIC) \cite{AbdulKhalek:2021gbh,AbdulKhalek:2022erw} states 
three-dimensional ``tomographic'' imaging of the proton as a major science goal.
Studies of the  deeply-virtual Compton scattering (DVCS) are an important part in this endeavour. 
This reaction gives access to the generalized parton distributions (GPDs)~\cite{Muller:1994ses,Ji:1996nm,Radyushkin:1997ki} 
that encode the information on the transverse position of quarks and gluons in the proton with dependence on their longitudinal momentum.
It will be measured with unprecedented precision and in a very broad kinematic range. 
The insights from these measurements can be fundamental and manifold. For instance, we want to find out whether quarks and gluons have similar 
distributions in transverse space. 
Form factors of the energy-momentum tensor can be accessed.
We will be able to study many related quantities --- the pressure distribution inside the proton, the helicity and angular momentum 
carried by quarks and gluons with different momentum fractions --- and address many other questions that are currently only tractable
within models or on the lattice.  

The QCD description of the DVCS is based on collinear factorization with GPDs as nonperturbative inputs and coefficient functions (CFs) which 
can be calculated order by order in perturbation theory. Since gluons do not have electric charge, to the leading order (LO), the whole
gluon GPD contribution is generated by mixing with the quark GPD and vanishes at the input scale, usually taken to be $Q_0^2=4$~GeV$^2$.
The next-to-leading order (NLO) corrections (one loop) were calculated for massless quarks and gluons in~\cite{Ji:1998xh} and for massive (e.g. charm) 
quarks in~\cite{Noritzsch:2003un}. The  next-to-next-to-leading order (NNLO) corrections (two loop) in conformal moments space
were estimated in~\cite{Kumericki:2006xx,Kumericki:2007sa} in a special renormalization scheme using constraints from conformal symmetry.  
For flavor-nonsinglet (NS) case, they were  calculated recently in momentum fraction space in \MS scheme 
in \cite{Braun:2020yib} and \cite{Braun:2021grd,Gao:2021iqq} for vector and axial-vector contributions, respectively.
In this letter we extend these results to flavor-singlet vector contributions that are crucial in the EIC energy range. 
We report on the calculation of NNLO flavor-singlet CFs for the Compton form factors (CFFs) $\f H$ and $\f E$
and an exploratory numerical analysis of their effect. Although the results are naturally dependent on the choice of 
model GPDs, the main conclusions are quite universal.
As a side result, we have re-derived and confirmed the expression for the NNLO flavor-nonsinglet CF, 
which was obtained in \cite{Braun:2020yib} using a different method.
 
The motivation for this work is twofold. On the one hand, the NNLO accuracy in DVCS is required for theoretical
consistency with the QCD studies of inclusive processes and also semi-inclusive reactions described in the 
transverse-momentum-dependent factorization framework where NNLO accuracy has become standard.  
More specifically, there are reasons to expect that the higher-order corrections to the gluon CFs are numerically large. 
It is well known, see e.g.~\cite{Freund:2001rk,Moutarde:2013qs}, although not emphasized strongly in the literature,
that the NLO (one loop) gluon CF has opposite sign compared to the leading quark contribution, 
leading to a strong cancellation between quarks and gluons at moderate $Q^2$.
At large photon virtualities this (formally subleading) negative gluon contribution is 
overcompensated by the positive contribution generated by the mixing with quarks, 
so that ultimately quark and gluon GPD contributions  add up.
This nontrivial interplay of LO and NLO continues through the whole range of accessible  momentum transfers at JLAB 12 and, in future, at EIC.
We find that the NNLO corrections follow the same pattern and are in all cases negative compared to the LO
(i.e. same sign as NLO). They are of the order of 20-30\% of the NLO gluon CF and significantly smaller for quarks, 
indicating that perturbation theory converges reasonably well. 
However, since all corrections are negative compared to the leading term, going over from NLO to NNLO the CFF $\f H$ is strongly reduced.
Our results reinforce the conclusions of \cite{Moutarde:2013qs} that the proper account of gluon contributions 
is crucial for a quantitative description of DVCS in the full energy range.  

%
\section{The calculation}
%
The DVCS scattering amplitude $\gamma^*(q) + N(p) \rightarrow \gamma(q') + N(p')$ 
is given by the nucleon (proton) matrix element of the time-ordered product of two electromagnetic currents
\begin{align}
T^{\mu \nu} & = i \int d^4x~e^{-iqx} \bra{p'} T \{ j^{\mu}(x) j^{\nu}(0) \} \ket{p}
\notag\\
&= - g_{\perp}^{\mu \nu} V + \epsilon_{\perp}^{\mu \nu} A + \text{ power corrections},
\end{align}
where $j^{\mu}(x) = \sum_q e_q \bar \psi_q(x) \gamma^{\mu} \psi_q(x)$ with the sum running over active quark flavors $q = u,d,s,...$ of electric charge $e_q$.  Transverse directions can be defined, e.g., as orthogonal to the scattering plane. (An ambiguity in the separation of longitudinal and transverse directions is a higher-twist effect, see \cite{Braun:2014sta,Guo:2021gru}.) 

The vector amplitude $V$ which is the subject of this work is usually parametrized in terms of two Compton form factors (CFFs)
\cite{Diehl:2003ny,Belitsky:2005qn}:
\begin{align}
V &=  \frac{1}{2P^+} \bar{u}(p') \biggl[\gamma^+ \mathcal{H}(\xi,Q,t)   +  \frac{i \sigma^{+\alpha}\Delta_{\alpha}}{2M}  
\mathcal{E}(\xi,Q,t)\biggr ]u(p)\,,
\end{align}
where $P_\mu = \tfrac12(p+p')_\mu$, $\Delta_\mu = (q'-q)_\mu$, $Q^2 = - q^2$  and $M$ is the nucleon mass.
The skewedness parameter $\xi$ to leading-twist accuracy can be chosen as~\cite{Belitsky:2005qn} 
\begin{align}
 \xi = \frac{x_B}{2-x_B} + \mathcal{O}(1/Q^2)\,,
\end{align}
where $x_B = Q^2/(2pq)$ is the usual Bjorken variable.

The CFFs can be factorized in terms of the GPDs,
\begin{align}
\mathcal{H}&= \sum_q \int_{-1}^1 \frac{dx}{\xi} C_q(x/\xi, \mu^2/Q^2, \alpha_s(\mu)) H_q(x,\xi,t,\mu)
\notag\\&\quad
 + \int_{-1}^1 \frac{dx}{\xi^2} C_g(x/\xi, \mu^2/Q^2, \alpha_s(\mu)) H_g(x,\xi,t,\mu),
\label{eq:CFF-H}
\end{align}
and similarly for $\mathcal{E}$, with the same coefficient functions (CFs). Here $\mu$ is the factorization scale and we tacitly take the renormalization scale to be same.
The quark and gluon GPDs are assumed to be normalized to the corresponding  parton distribution functions (PDFs) in the forward limit
\cite{Diehl:2003ny,Belitsky:2005qn} 
\begin{align}
  H_q(x,0,0,\mu) = q(x,\mu)\,, \qquad
  H_g(x,0,0,\mu) = x g(x,\mu)\,.
\label{eq:forward}
\end{align}
The coefficients $1/\xi$ and $1/\xi^2$ in (\ref{eq:CFF-H}) in front of the quark and gluon contributions 
ensure that the corresponding CFs $C_q$ and $C_g$ depend on the ratio $x/\xi$ only.
The CFs are real functions for $|x/\xi| <1$ and can be continued analytically to the $|x/\xi| > 1$ region by the substitution
$\xi \to \xi -i\epsilon$~\cite{Ji:1996nm,Radyushkin:1997ki}.

The CFs can be calculated order-by-order in perturbation theory 
\begin{align}
C_q &= C_q^{(0)} + a_s C_q^{(1)} + a_s^2 C_q^{(2)} + \f O(a_s^3),
\notag\\
C_g &= a_s C_g^{(1)} + a_s^2 C_g^{(2)} + \f O(a_s^3),
\label{eq:CF1}
\end{align}
where $a_s = \alpha_s/(4\pi)$ and  in our normalization
\begin{align}
  C_q^{(0)} &= e_q^2 \left(\frac{\xi}{\xi-x -i\epsilon} - \frac{\xi}{\xi+x-i\epsilon}\right).
\label{eq:CF-LO}
\end{align}
The NLO (one-loop) CFs  are available from~\cite{Ji:1998xh,Noritzsch:2003un}. The NNLO (two-loop) CFs present our 
main result. The answer can be decomposed in contributions of different color structures and, for quarks, 
in flavor-nonsinglet (NS) and pure-flavor singlet (PS) contributions with different dependence on electromagnetic 
charges: 
\begin{align}
C_q^{(2)} &= \frac{1}{2z(1-z)}  \biggl[ e_q^2 C_F \Big( C_F \text C_{\text{NS}}^{(F)}+ C_A \text C_{\text{NS}}^{(A)} 
+ \beta_0 \text C_{\text{NS}}^{(\beta_0)} \Big) 
\notag\\&\quad 
+ \Big ( \sum_{q'} e_{q'}^2 \Big ) T_F C_F \mathrm{C}_{\text{PS}} \biggr] \label{eq: CF2quark},
\\
C_g^{(2)} &= \frac{\Big ( \sum_q e_q^2 \Big )}{4z^2(1-z)^2} T_F  \Big ( C_F \text C_{g}^{(F)} + C_A \text C_{g}^{(A)} \Big ),
\label{eq:CF2}
\end{align}
where $z = \tfrac12(1-x/\xi)$ is the argument of all ${\rm C}_{X}^{(Y)}$ implicitly. All quark CFs are antisymmetric and gluon CFs symmetric under the interchange $x\leftrightarrow -x$ alias 
$z\leftrightarrow 1-z$.

The calculation was performed using computer algebra techniques. 
The $\sim 150$ contributing Feynman diagrams were generated using \texttt{qgraf} \cite{Nogueira:1991ex}
and projected on to the appropriate  Dirac and Lorentz structures using \texttt{FORM} \cite{Vermaseren:2000nd} and in-house routines.
The resulting set of scalar integrals was reduced to 
12 master integrals making use of the integration-by-parts relations, performed with \texttt{FIRE} \cite{Smirnov:2019qkx}.
Fortunately, no new master integrals appeared in the singlet CFs compared to the non-singlet case. 
These integrals have been calculated recently in \cite{Gao:2021iqq} and are partially given in \cite{Gao:2021beo}. The divergent $1/\epsilon$ terms in the resulting expressions
are all removed when expressing the results in terms of renormalized GPDs and the renormalized coupling, which provides 
a strong check of the calculation. On this way also finite contributions appear due to 
 convolutions of the $\mathcal{O}(\epsilon) $ terms in the one-loop CFs with the one-loop evolution kernels. 
 These convolutions were performed in both momentum and position space with the latter using \texttt{HyperInt} software package~\cite{Panzer:2014caa}.
The final manipulations and the numerical analysis reported below were done using the \texttt{HPL}
package~\cite{Maitre:2005uu} for handling harmonic polylogarithms \cite{Remiddi:1999ew}.

Complete analytic expressions for the CFs $C_{\text{NS}}$, $C_{\text{PS}}$, $C_{{g}}$ up to two loops 
are given in the Supplemental Materials  and collected in the ancillary file \texttt{CF.m} in \texttt{Mathematica} format. 
Our results for the two-loop flavor-nonsinglet CFs $\mathrm{C}_{\text{NS}}^{(F)}$, $\mathrm{C}_{\text{NS}}^{(A)}$
and $\mathrm{C}_{\text{NS}}^{(\beta_0)}$ agree with the corresponding results in Ref.~\cite{Braun:2020yib} obtained using a different, 
conformal symmetry based  approach.

We observe (see Supplemental Materials) that the quark CF becomes smaller with the inclusion of
higher-order effects, the reason 
being that the one-loop and two-loop corrections both have
opposite sign as compared to the LO term \eqref{eq:CF-LO}
apart from a very narrow strip around $x/\xi=1$. 
The gluon CF becomes significantly larger by adding the two-loop effects, 
and has negative sign. Thus the quark and gluon contributions to the CFFs 
at $\mu^2 = Q^2$ have opposite sign, leading to a strong cancellation 
in the sum.

\begin{figure*}[t]
\includegraphics[width=0.32\textwidth]{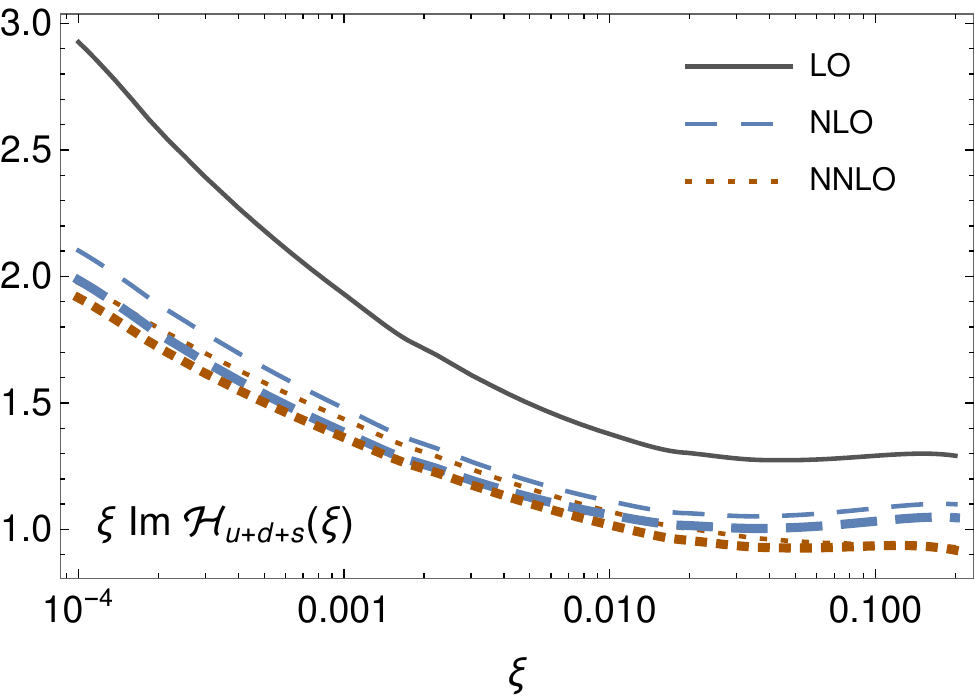}
\includegraphics[width=0.325\textwidth]{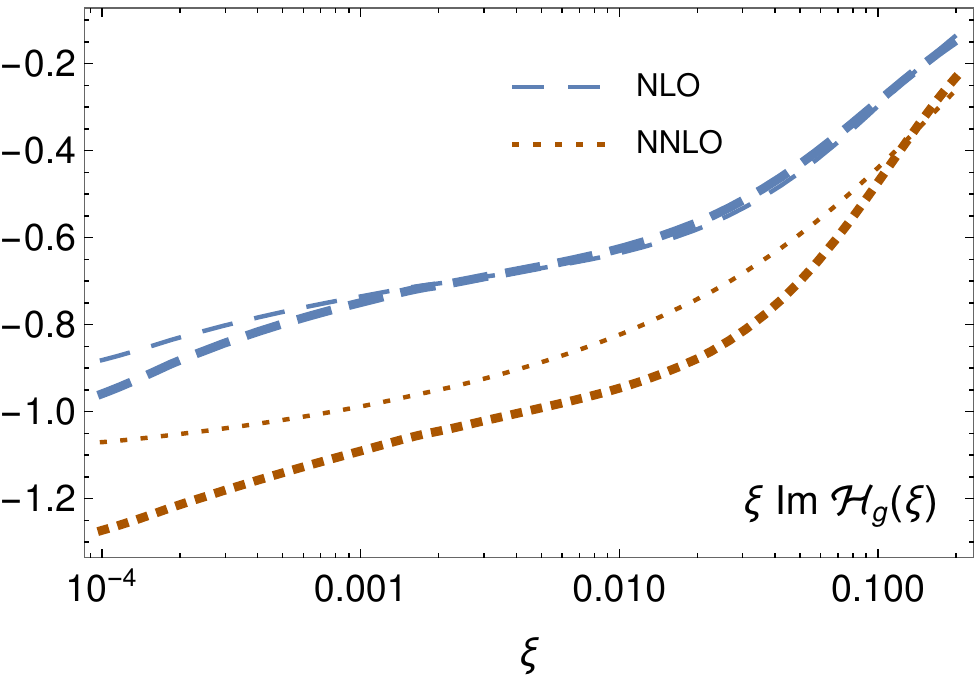}
\includegraphics[width=0.32\textwidth]{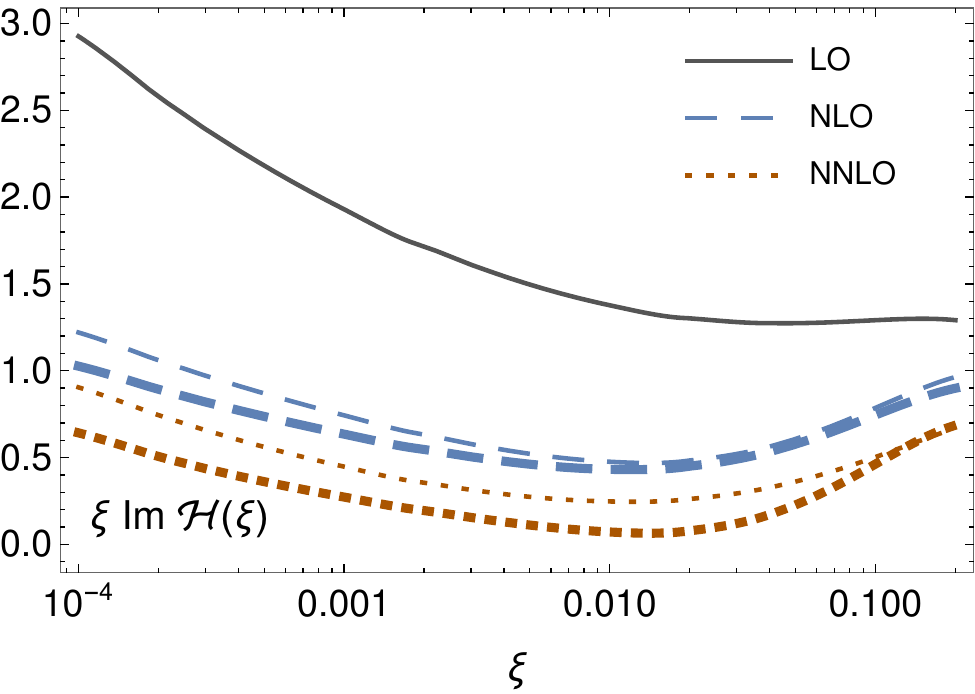}
\includegraphics[width=0.32\textwidth]{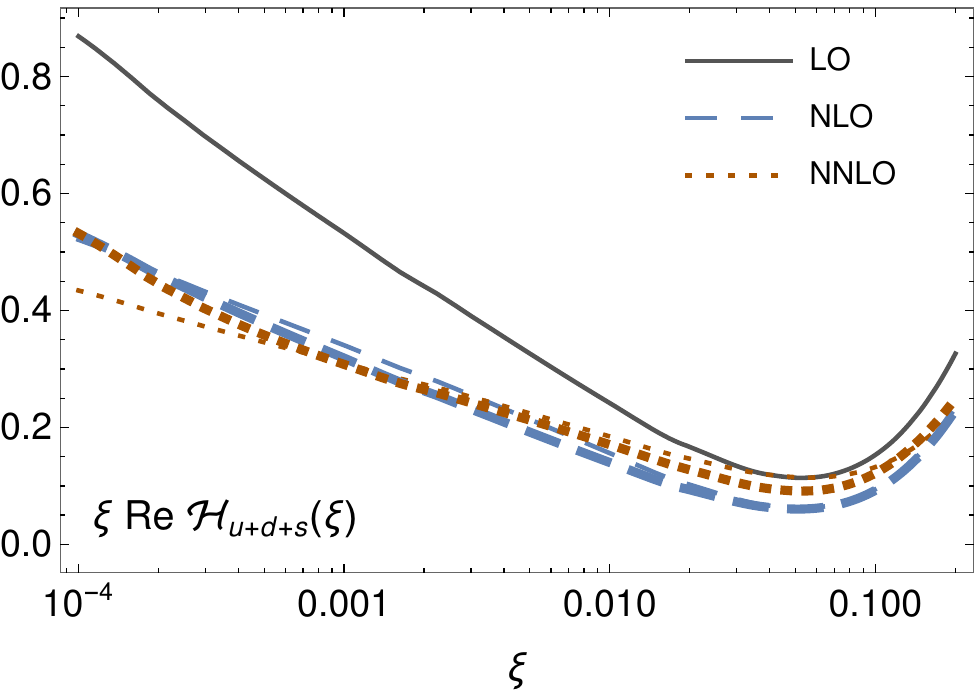}
\includegraphics[width=0.325\textwidth]{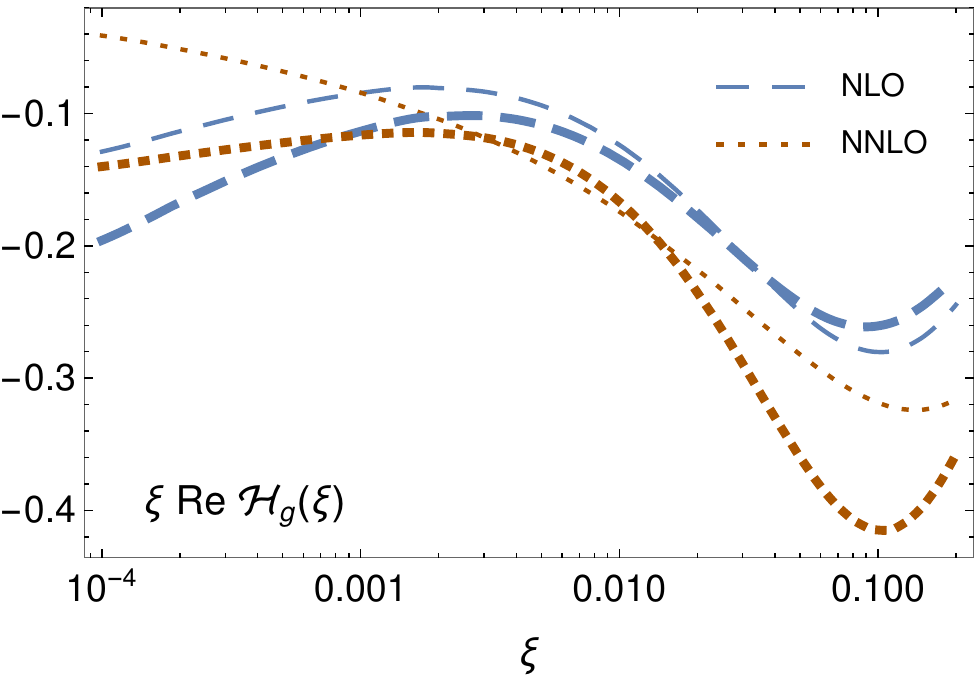}
\includegraphics[width=0.32\textwidth]{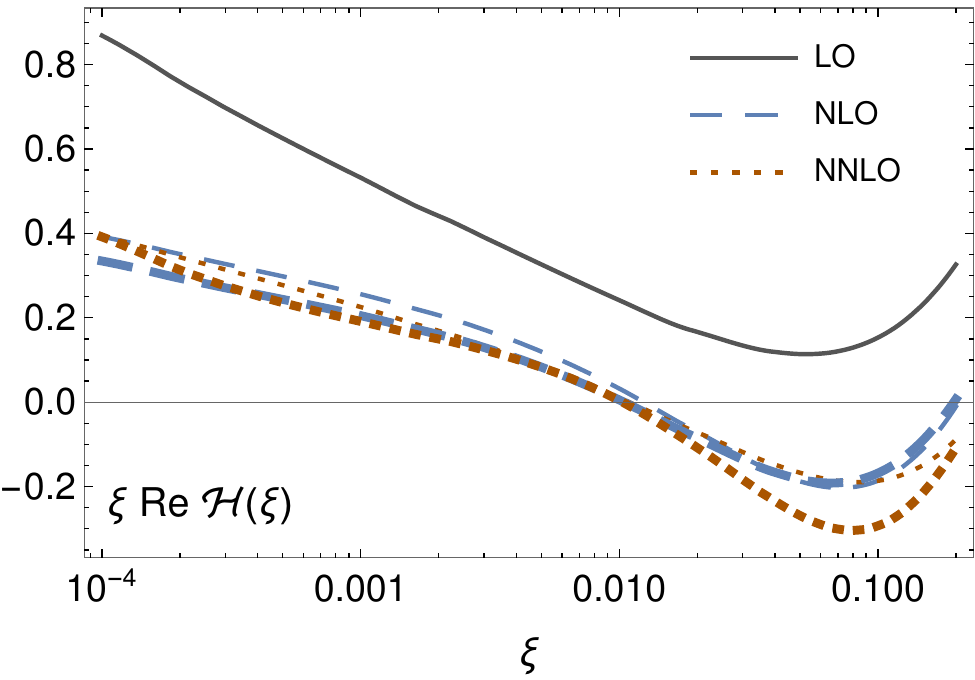}
\caption{Real and imaginary parts of the CFF $\f H$ as a function of $\xi$ at $\mu^2 = Q^2 = 4\text{ GeV}^2$ and $t = -0.1\text{ GeV}^2$ for the GPDs normalized to HERAPDF20 (thin lines) and ABMP16 (thick lines, NLO and NNLO only) PDF sets at the appropriate order
in perturbation theory: 
Solid lines: LO (black), short dashes: NLO (blue), long dashes: NNLO (orange).}
\label{figH4GeVnew}
\end{figure*}

The most singular contributions at $z \to 0$ ($x\to\xi$) increase by two powers of the logarithm for each higher power of the coupling. We obtain
\begin{align}
 C_q &\simeq \frac{e_q^2}{2 z}\Big(1 + C_F a_s \ln^2 z + \frac12 C^2_F a^2_s \ln^4 z + \ldots\Big),
\notag\\
 C_g &\simeq  \frac{\sum_q e_q^2}{2z} T_F  a_s \ln z \Big(1 + \frac{1}{6}(C_A+5C_F) a_s \ln^2 z + \ldots \Big).
\label{eq:singularity}
\end{align}
The ``pure singlet'' quark contribution $\mathrm{C}_{\text{PS}}$ \eqref{eq:CF2} does not contribute to this asymptotic behavior at NNLO so that the double-logarithmic 
asymptotics for $C_q$ is given by the flavor-nonsinglet contribution alone.
Our expression in the first line in \eqref{eq:singularity} coincides with the one obtained in \cite{Braun:2020yib}, 
but does not agree with the resummation formula suggested in~\cite{Altinoluk:2012nt}. 

%
\section{Numerical analysis}
%

For the illustration of the numerical impact of the NNLO corrections on the CFF $\mathcal{H}(\xi,Q^2, t)$, we will use models for the GPDs based 
on the conventional double-distributions parametrization~\cite{Radyushkin:1997ki,Radyushkin:1998bz}
\begin{align}
H_i(x,\xi, t) &= \iint\limits_{|\alpha| + |\beta| \leq 1} \!\!\!d\alpha\, d\beta\, f_i(\beta, \alpha, t) \delta(x - \beta - \xi \alpha)
\notag\\ &\quad
+ D_i(x/\xi,t) \theta(\xi^2-x^2)\, ,
\end{align}
where the subscript $i$ denotes the flavor ($val$ for valence quarks, $sea$ for sea quarks and $g$ for gluons). 
We neglect the $D$-term \cite{Polyakov:1999gs} in what follows, as it 
is very poorly known and does not contribute to the dominant imaginary part of the CFF $\mathcal{H}$ at high energies.  
We also neglect the  $c$-quark contribution which, according to the analysis in~\cite{Noritzsch:2003un}, 
is very small at such scales.   

The GPDs are constructed  using the standard ansatz \cite{Musatov:1999xp}
\begin{align}
f_i(\beta,\alpha,t) &= g_i(\beta,t) h_i(\beta) \frac{\Gamma(2n_i + 2)}{2^{2n_i+1} \Gamma(n_i+1)^2} 
\notag\\&\quad 
\times \frac{[(1-|\beta|)^2 - \alpha^2]^{n_i}}{(1-|\beta|)^{2n_i+1} },
\end{align}
with $n_{val} =1$ and $ n_{g}= n_{sea}=2 $. The form of the $t$-dependence is inspired by the 
Regge calculus~\cite{Goloskokov:2006hr,Kroll:2012sm}
\begin{align}
  g_i(\beta,t) &= e^{b_i t} |\beta|^{-\alpha'_i t}\,,
\label{gi}
\end{align}
with parameters specified in Ref.~\cite{Goloskokov:2006hr}. 
The functions $h_i(\beta)$ are related to the corresponding PDFs through the normalization condition (\ref{eq:forward}):
\begin{align*}
h_g(\beta) &= |\beta| g(|\beta|)\,, 
\\
h_{sea}(\beta) &= q_{sea}(|\beta|) \text{sign}(\beta)\,,
\\
h_{val}(\beta) &= q_{val}{(\beta)} \theta(\beta)\,.
\end{align*}

It is well known that the PDFs from global fits depend rather strongly on the order in perturbation theory.
We use HERAPDF20~\cite{H1:2015ubc}  LO/NLO/NNLO PDFs for the calculation of the CFF 
$\mathcal{H}(\xi,Q,t)$ including
LO/NLO/NNLO CFs, respectively. The results are shown in Fig.~\ref{figH4GeVnew} by thin curves.
We also show by thick curves the NLO and NNLO results using the ABMP16  \cite{Alekhin:2017kpj,Alekhin:2018pai}
PDF sets for comparison. Quark contributions ($u+d+s$), gluon contributions, and the total result for  
$\mathcal{H}(\xi,Q,t)$ are shown on the left, middle, and right panels in Fig.~\ref{figH4GeVnew}, respectively. We have chosen 
the momentum transfer $t=-0.1\,\text{GeV}^2$ as a representative value for these plots.

One sees that the NNLO corrections to the quark contributions are rather small or, rather,  partially compensated 
by refitting the GPD model to the NNLO PDFs in the forward limit. However, the NNLO contribution is large for gluons 
with $\sim (20-30)\%$ for the imaginary part. Since the gluon contribution to $\text{Im}\mathcal{H}(\xi,t)$ at 
$\mu^2=Q^2$ is large and negative, the total NNLO correction to the imaginary part of the CFF at the input scale 
appears to be very large, 
up to a factor of two compared to NLO at $\xi \sim 0.01$.
Note that this cancellation becomes stronger for increasing values of $t$ as by assumption \eqref{gi}
the flavor-singlet contribution has a smaller slope. 
We believe that these conclusions are quite universal 
although the results presented in Fig.~\ref{figH4GeVnew} are naturally dependent on the used GPD model which may be too simplistic.

Next, we consider the $Q^2$ dependence. Lacking the three-loop evolution equations for the flavor-singlet GPDs, 
a complete NNLO calculation is not possible at this time. 
A sufficiently flexible numerical code for the two-loop evolution equations is only available in 
conformal moments space \cite{Kumericki:2007sa}%
\footnote{For the one-loop evolution, two highly efficient codes exist \cite{Vinnikov:2006xw,Bertone:2022frx} 
and are implemented on the PARTONS platform \cite{Berthou:2015oaw}.}.
For this reason we decided not to use the renormalization group improved expressions in what follows, but implement the $Q^2$ 
dependence as encoded in the coefficient functions up to fixed order, $\mathcal{O}(\alpha_s^2)$. In this way the GPDs are kept 
at $\mu^2=4\, \text{GeV}^2$ and $Q^2$ dependence at LO
is introduced by taking into account terms $\mathcal{O}(\{1, a_s L, a^2_s L^2\})$ with $L=\ln\mu^2/Q^2$ in the CFs, 
corresponding to the first two iterations of the LO evolution equation.  
The NLO results are obtained by adding terms $\mathcal{O}(\{a_s, a^2_s L\})$ and at NNLO
we use complete calculated CFs, adding terms $\mathcal{O}(a^2_s)$ independent of $L$. 
This approximation is sufficient for 
our purposes as the accessible $Q^2$ range in the ongoing and planned measurements is not large.

\begin{figure*}[t]
\includegraphics[width=0.32\textwidth]{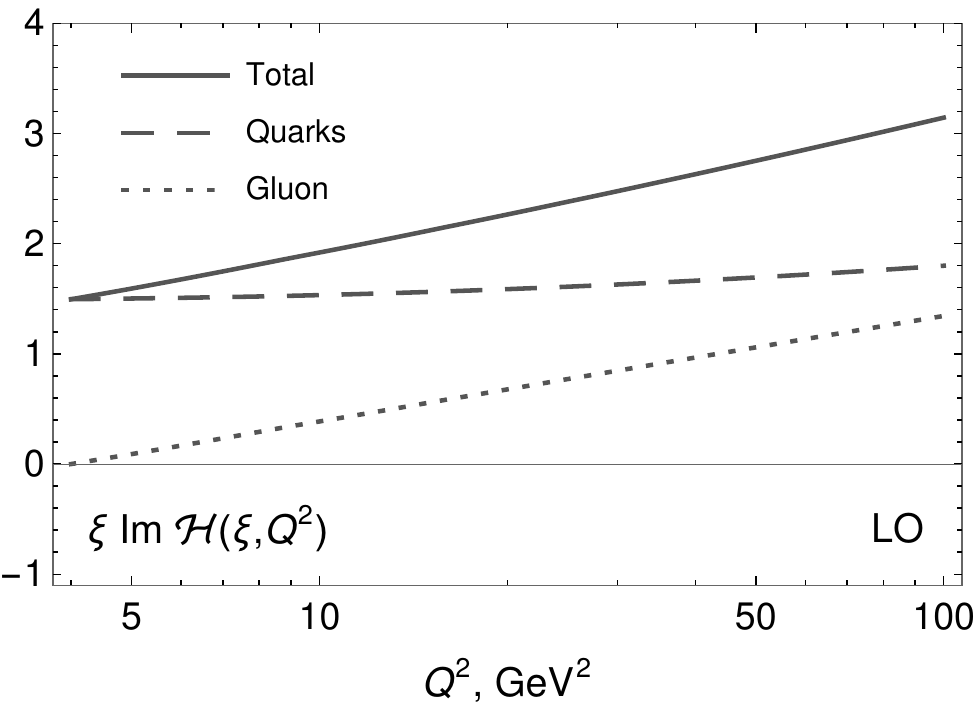}
\includegraphics[width=0.325\textwidth]{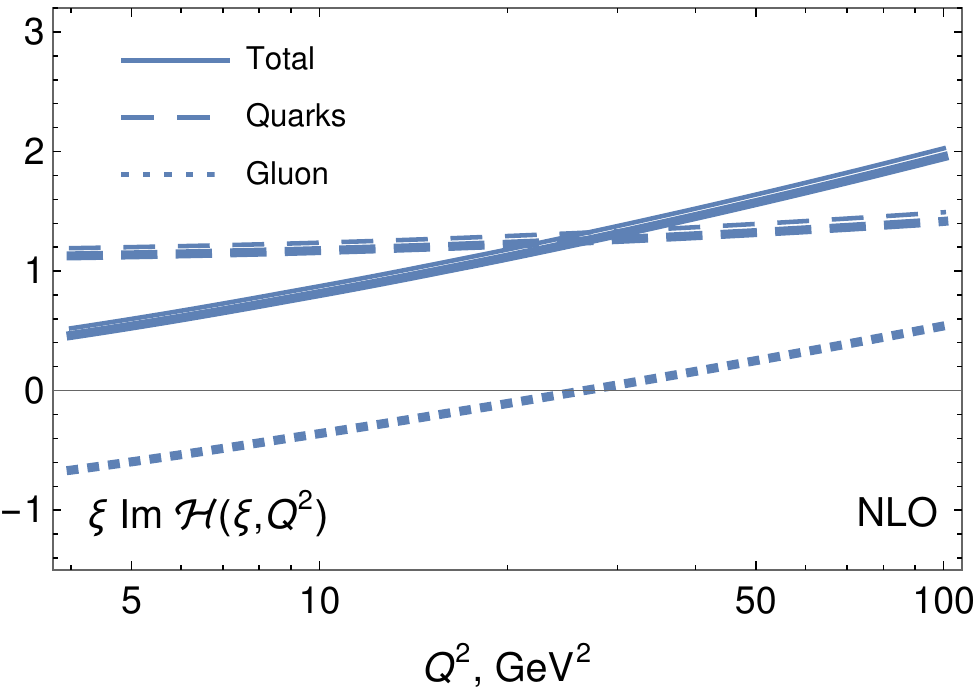}
\includegraphics[width=0.32\textwidth]{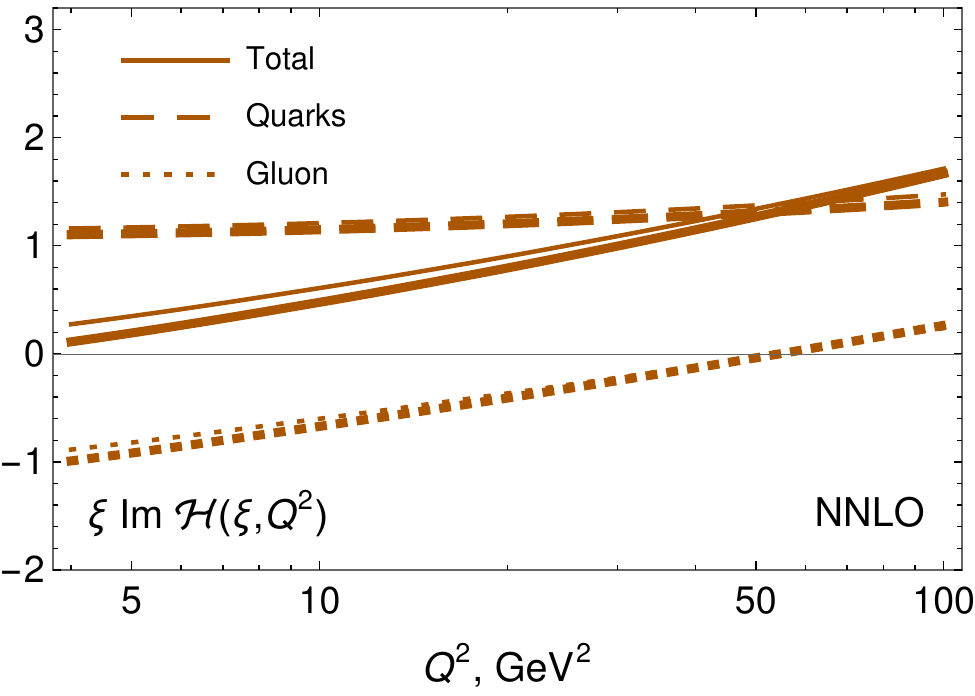}
\includegraphics[width=0.32\textwidth]{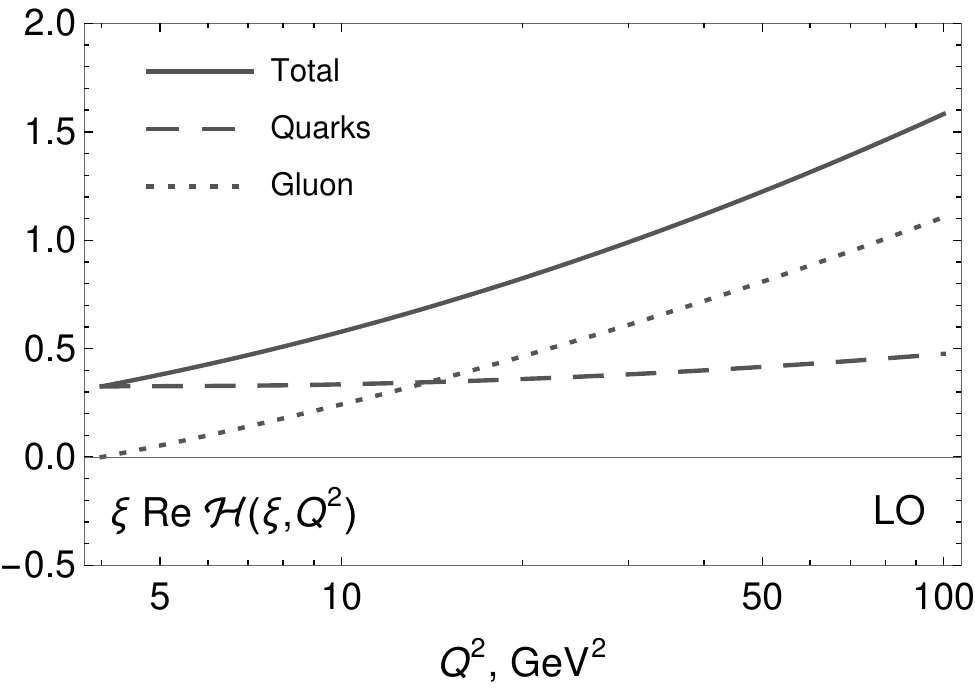}
\includegraphics[width=0.325\textwidth]{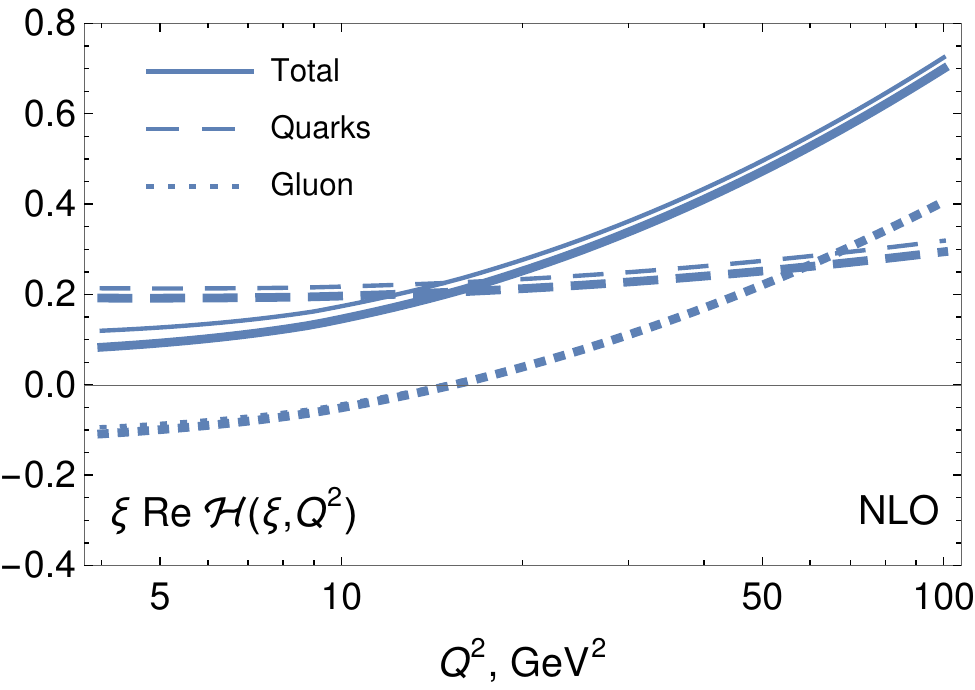}
\includegraphics[width=0.32\textwidth]{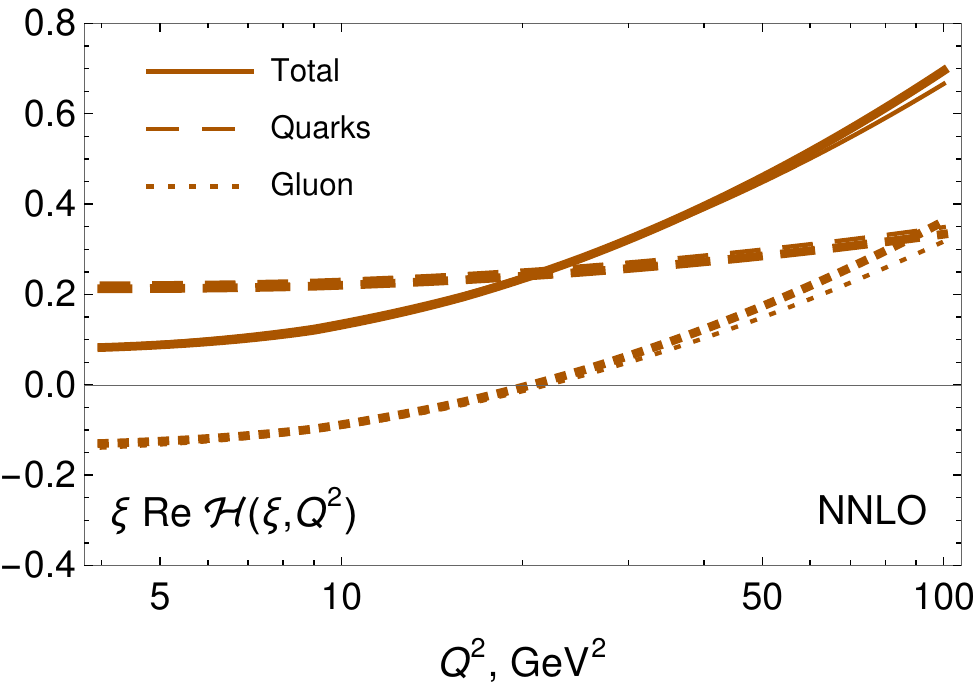}
\caption{The $Q^2$ dependence of the imaginary (upper panels) and real (low panels) parts of the CFF $\f H$ for $\xi = 0.005$
and $t = -0.1\text{ GeV}^2$ for the GPDs at the input scale $\mu^2 = 4\text{ GeV}^2$ normalized to HERAPDF20 (thin lines) and ABMP16 
 PDF sets at the appropriate order in perturbation theory. The quark and gluon contributions are
shown by dashed and dotted curves, respectively. Their sum is shown by the solid lines.}
\label{figHQ2}
\end{figure*}

The results for a particular choice $\xi =0.005$ are shown in Fig.~\ref{figHQ2}.
 At LO (left panels), the gluon GPD contribution (dots) vanishes at the input scale
 $Q^2 = \mu^2 = 4\text{ GeV}^2$ and is generated at higher scales due to mixing with quarks. The $Q^2$ dependence of the quark contribution 
shown by dashed curves increases very slowly. The sum of the quark and gluon contribution (solid curves) increases by roughly a factor two  
for $4\, {\rm GeV}^2 \leq Q^2 \leq 100\,\text{GeV}^2$. This increase is driven almost entirely by the gluons.

At NLO, both quark and gluon CFs receive negative $\mathcal{O}{(\alpha_s)}$  corrections so that the quark contribution decreses slightly
and the gluon contribution becomes large and negative.  As the result, the CFF $\f H$ is strongly reduced. 
With increasing $Q^2$, this (formally subleading) negative correction 
is gradually compensated by a positive contribution from quark-gluon mixing so that the total gluon contribution 
changes sign at $Q^2 \sim 25$~GeV$^2$.

The main effect of the NNLO corrections $\mathcal{O}{(\alpha^2_s)}$ calculated in this work is a large additional negative gluon contribution
to the imaginary part of the CFF. As the result, the cancellation between the quark and gluon contributions at moderate $Q^2 < 20\,\text{GeV}^2$ 
becomes stronger and the CFF $\f H$ is further reduced --- by  a factor of two or more at $\xi =0.005$, depending on the GPD model.   

Conversely, if the quark and gluon GPDs are extracted by fitting the QCD theory predictions to the experimental data on the CFF,
the results are likely going to be very unstable against changes in the order of perturbation theory in the analysis.

%
\section{Conclusions}
%

To summarize, in anticipation of the high-precision experimental data on DVCS in a broad kinematic range 
from JLAB 12 and the Electron-Ion Collider,
we have calculated the two-loop, (NNLO) CFs associated with the dominant 
Compton form factors $\f H$ and $\f E$ at large energies. 
We find that the NNLO correction to the gluon contribution to the imaginary part of  $\f H$ is significant so that
the cancellation between quark and gluons at moderate $Q^2$ (that is already present at NLO) becomes more pronounced.
To tame this instability, a further analysis is needed both in theory and phenomenology. 
From the theory side, in particular the three-loop evolution equations for flavor-singlet GPDs
are missing and will be necessary to quantify the remaining factorization scale dependence at NNLO.
Also an estimate of the three-loop (N$^3$LO) CF is obtainable by calculating the two-loop diagrams with 
an additional fermion-bubble insertion. Both tasks, however, go beyond the scope of this letter.

%
\begin{acknowledgments}
\section*{Acknowledgements}
We thank  V.~Bertone, K.~Kumericki, and P.~Sznajder for correspondence on the topic of the 
scale dependence.
This work was supported in part by the Research Unit FOR2926 and 
the Collaborative Research Center TRR110/2 funded by 
the Deutsche Forschungsgemeinschaft (DFG, German Research Foundation) under grants No. 409651613 and 
No. 196253076, respectively.     
\end{acknowledgments}
%



\begin{thebibliography}{37}%
\makeatletter
\providecommand \@ifxundefined [1]{%
 \@ifx{#1\undefined}
}%
\providecommand \@ifnum [1]{%
 \ifnum #1\expandafter \@firstoftwo
 \else \expandafter \@secondoftwo
 \fi
}%
\providecommand \@ifx [1]{%
 \ifx #1\expandafter \@firstoftwo
 \else \expandafter \@secondoftwo
 \fi
}%
\providecommand \natexlab [1]{#1}%
\providecommand \enquote  [1]{``#1''}%
\providecommand \bibnamefont  [1]{#1}%
\providecommand \bibfnamefont [1]{#1}%
\providecommand \citenamefont [1]{#1}%
\providecommand \href@noop [0]{\@secondoftwo}%
\providecommand \href [0]{\begingroup \@sanitize@url \@href}%
\providecommand \@href[1]{\@@startlink{#1}\@@href}%
\providecommand \@@href[1]{\endgroup#1\@@endlink}%
\providecommand \@sanitize@url [0]{\catcode `\\12\catcode `\$12\catcode
  `\&12\catcode `\#12\catcode `\^12\catcode `\_12\catcode `\%12\relax}%
\providecommand \@@startlink[1]{}%
\providecommand \@@endlink[0]{}%
\providecommand \url  [0]{\begingroup\@sanitize@url \@url }%
\providecommand \@url [1]{\endgroup\@href {#1}{\urlprefix }}%
\providecommand \urlprefix  [0]{URL }%
\providecommand \Eprint [0]{\href }%
\providecommand \doibase [0]{http://dx.doi.org/}%
\providecommand \selectlanguage [0]{\@gobble}%
\providecommand \bibinfo  [0]{\@secondoftwo}%
\providecommand \bibfield  [0]{\@secondoftwo}%
\providecommand \translation [1]{[#1]}%
\providecommand \BibitemOpen [0]{}%
\providecommand \bibitemStop [0]{}%
\providecommand \bibitemNoStop [0]{.\EOS\space}%
\providecommand \EOS [0]{\spacefactor3000\relax}%
\providecommand \BibitemShut  [1]{\csname bibitem#1\endcsname}%
\let\auto@bib@innerbib\@empty
\bibitem [{\citenamefont {Abdul~Khalek}\ \emph {et~al.}(2021)\citenamefont
  {Abdul~Khalek} \emph {et~al.}}]{AbdulKhalek:2021gbh}%
  \BibitemOpen
  \bibfield  {author} {\bibinfo {author} {\bibfnamefont {R.}~\bibnamefont
  {Abdul~Khalek}} \emph {et~al.},\ }\href@noop {} {\  (\bibinfo {year}
  {2021})},\ \Eprint {http://arxiv.org/abs/2103.05419} {arXiv:2103.05419
  [physics.ins-det]} \BibitemShut {NoStop}%
\bibitem [{\citenamefont {Abdul~Khalek}\ \emph {et~al.}(2022)\citenamefont
  {Abdul~Khalek} \emph {et~al.}}]{AbdulKhalek:2022erw}%
  \BibitemOpen
  \bibfield  {author} {\bibinfo {author} {\bibfnamefont {R.}~\bibnamefont
  {Abdul~Khalek}} \emph {et~al.},\ }in\ \href@noop {} {\emph {\bibinfo
  {booktitle} {{2022 Snowmass Summer Study}}}}\ (\bibinfo {year} {2022})\
  \Eprint {http://arxiv.org/abs/2203.13199} {arXiv:2203.13199 [hep-ph]}
  \BibitemShut {NoStop}%
\bibitem [{\citenamefont {M\"uller}\ \emph {et~al.}(1994)\citenamefont
  {M\"uller}, \citenamefont {Robaschik}, \citenamefont {Geyer}, \citenamefont
  {Dittes},\ and\ \citenamefont {Ho\v{r}ej\v{s}i}}]{Muller:1994ses}%
  \BibitemOpen
  \bibfield  {author} {\bibinfo {author} {\bibfnamefont {D.}~\bibnamefont
  {M\"uller}}, \bibinfo {author} {\bibfnamefont {D.}~\bibnamefont {Robaschik}},
  \bibinfo {author} {\bibfnamefont {B.}~\bibnamefont {Geyer}}, \bibinfo
  {author} {\bibfnamefont {F.~M.}\ \bibnamefont {Dittes}}, \ and\ \bibinfo
  {author} {\bibfnamefont {J.}~\bibnamefont {Ho\v{r}ej\v{s}i}},\ }\href
  {\doibase 10.1002/prop.2190420202} {\bibfield  {journal} {\bibinfo  {journal}
  {Fortsch. Phys.}\ }\textbf {\bibinfo {volume} {42}},\ \bibinfo {pages} {101}
  (\bibinfo {year} {1994})},\ \Eprint {http://arxiv.org/abs/hep-ph/9812448}
  {arXiv:hep-ph/9812448} \BibitemShut {NoStop}%
\bibitem [{\citenamefont {Ji}(1997)}]{Ji:1996nm}%
  \BibitemOpen
  \bibfield  {author} {\bibinfo {author} {\bibfnamefont {X.-D.}\ \bibnamefont
  {Ji}},\ }\href {\doibase 10.1103/PhysRevD.55.7114} {\bibfield  {journal}
  {\bibinfo  {journal} {Phys. Rev. D}\ }\textbf {\bibinfo {volume} {55}},\
  \bibinfo {pages} {7114} (\bibinfo {year} {1997})},\ \Eprint
  {http://arxiv.org/abs/hep-ph/9609381} {arXiv:hep-ph/9609381} \BibitemShut
  {NoStop}%
\bibitem [{\citenamefont {Radyushkin}(1997)}]{Radyushkin:1997ki}%
  \BibitemOpen
  \bibfield  {author} {\bibinfo {author} {\bibfnamefont {A.~V.}\ \bibnamefont
  {Radyushkin}},\ }\href {\doibase 10.1103/PhysRevD.56.5524} {\bibfield
  {journal} {\bibinfo  {journal} {Phys. Rev. D}\ }\textbf {\bibinfo {volume}
  {56}},\ \bibinfo {pages} {5524} (\bibinfo {year} {1997})},\ \Eprint
  {http://arxiv.org/abs/hep-ph/9704207} {arXiv:hep-ph/9704207} \BibitemShut
  {NoStop}%
\bibitem [{\citenamefont {Ji}\ and\ \citenamefont {Osborne}(1998)}]{Ji:1998xh}%
  \BibitemOpen
  \bibfield  {author} {\bibinfo {author} {\bibfnamefont {X.-D.}\ \bibnamefont
  {Ji}}\ and\ \bibinfo {author} {\bibfnamefont {J.}~\bibnamefont {Osborne}},\
  }\href {\doibase 10.1103/PhysRevD.58.094018} {\bibfield  {journal} {\bibinfo
  {journal} {Phys. Rev. D}\ }\textbf {\bibinfo {volume} {58}},\ \bibinfo
  {pages} {094018} (\bibinfo {year} {1998})},\ \Eprint
  {http://arxiv.org/abs/hep-ph/9801260} {arXiv:hep-ph/9801260} \BibitemShut
  {NoStop}%
\bibitem [{\citenamefont {Noritzsch}(2004)}]{Noritzsch:2003un}%
  \BibitemOpen
  \bibfield  {author} {\bibinfo {author} {\bibfnamefont {J.~D.}\ \bibnamefont
  {Noritzsch}},\ }\href {\doibase 10.1103/PhysRevD.69.094016} {\bibfield
  {journal} {\bibinfo  {journal} {Phys. Rev. D}\ }\textbf {\bibinfo {volume}
  {69}},\ \bibinfo {pages} {094016} (\bibinfo {year} {2004})},\ \Eprint
  {http://arxiv.org/abs/hep-ph/0312137} {arXiv:hep-ph/0312137} \BibitemShut
  {NoStop}%
\bibitem [{\citenamefont {Kumericki}\ \emph {et~al.}(2007)\citenamefont
  {Kumericki}, \citenamefont {Mueller}, \citenamefont {Passek-Kumericki},\ and\
  \citenamefont {Schafer}}]{Kumericki:2006xx}%
  \BibitemOpen
  \bibfield  {author} {\bibinfo {author} {\bibfnamefont {K.}~\bibnamefont
  {Kumericki}}, \bibinfo {author} {\bibfnamefont {D.}~\bibnamefont {Mueller}},
  \bibinfo {author} {\bibfnamefont {K.}~\bibnamefont {Passek-Kumericki}}, \
  and\ \bibinfo {author} {\bibfnamefont {A.}~\bibnamefont {Schafer}},\ }\href
  {\doibase 10.1016/j.physletb.2007.02.071} {\bibfield  {journal} {\bibinfo
  {journal} {Phys. Lett. B}\ }\textbf {\bibinfo {volume} {648}},\ \bibinfo
  {pages} {186} (\bibinfo {year} {2007})},\ \Eprint
  {http://arxiv.org/abs/hep-ph/0605237} {arXiv:hep-ph/0605237} \BibitemShut
  {NoStop}%
\bibitem [{\citenamefont {Kumericki}\ \emph {et~al.}(2008)\citenamefont
  {Kumericki}, \citenamefont {Mueller},\ and\ \citenamefont
  {Passek-Kumericki}}]{Kumericki:2007sa}%
  \BibitemOpen
  \bibfield  {author} {\bibinfo {author} {\bibfnamefont {K.}~\bibnamefont
  {Kumericki}}, \bibinfo {author} {\bibfnamefont {D.}~\bibnamefont {Mueller}},
  \ and\ \bibinfo {author} {\bibfnamefont {K.}~\bibnamefont
  {Passek-Kumericki}},\ }\href {\doibase 10.1016/j.nuclphysb.2007.10.029}
  {\bibfield  {journal} {\bibinfo  {journal} {Nucl. Phys. B}\ }\textbf
  {\bibinfo {volume} {794}},\ \bibinfo {pages} {244} (\bibinfo {year}
  {2008})},\ \Eprint {http://arxiv.org/abs/hep-ph/0703179}
  {arXiv:hep-ph/0703179} \BibitemShut {NoStop}%
\bibitem [{\citenamefont {Braun}\ \emph {et~al.}(2020)\citenamefont {Braun},
  \citenamefont {Manashov}, \citenamefont {Moch},\ and\ \citenamefont
  {Schoenleber}}]{Braun:2020yib}%
  \BibitemOpen
  \bibfield  {author} {\bibinfo {author} {\bibfnamefont {V.~M.}\ \bibnamefont
  {Braun}}, \bibinfo {author} {\bibfnamefont {A.~N.}\ \bibnamefont {Manashov}},
  \bibinfo {author} {\bibfnamefont {S.}~\bibnamefont {Moch}}, \ and\ \bibinfo
  {author} {\bibfnamefont {J.}~\bibnamefont {Schoenleber}},\ }\href {\doibase
  10.1007/JHEP09(2020)117} {\bibfield  {journal} {\bibinfo  {journal} {JHEP}\
  }\textbf {\bibinfo {volume} {09}},\ \bibinfo {pages} {117} (\bibinfo {year}
  {2020})},\ \Eprint {http://arxiv.org/abs/2007.06348} {arXiv:2007.06348
  [hep-ph]} \BibitemShut {NoStop}%
\bibitem [{\citenamefont {Braun}\ \emph {et~al.}(2021)\citenamefont {Braun},
  \citenamefont {Manashov}, \citenamefont {Moch},\ and\ \citenamefont
  {Schoenleber}}]{Braun:2021grd}%
  \BibitemOpen
  \bibfield  {author} {\bibinfo {author} {\bibfnamefont {V.~M.}\ \bibnamefont
  {Braun}}, \bibinfo {author} {\bibfnamefont {A.~N.}\ \bibnamefont {Manashov}},
  \bibinfo {author} {\bibfnamefont {S.}~\bibnamefont {Moch}}, \ and\ \bibinfo
  {author} {\bibfnamefont {J.}~\bibnamefont {Schoenleber}},\ }\href {\doibase
  10.1103/PhysRevD.104.094007} {\bibfield  {journal} {\bibinfo  {journal}
  {Phys. Rev. D}\ }\textbf {\bibinfo {volume} {104}},\ \bibinfo {pages}
  {094007} (\bibinfo {year} {2021})},\ \Eprint
  {http://arxiv.org/abs/2106.01437} {arXiv:2106.01437 [hep-ph]} \BibitemShut
  {NoStop}%
\bibitem [{\citenamefont {Gao}\ \emph {et~al.}(2022)\citenamefont {Gao},
  \citenamefont {Huber}, \citenamefont {Ji},\ and\ \citenamefont
  {Wang}}]{Gao:2021iqq}%
  \BibitemOpen
  \bibfield  {author} {\bibinfo {author} {\bibfnamefont {J.}~\bibnamefont
  {Gao}}, \bibinfo {author} {\bibfnamefont {T.}~\bibnamefont {Huber}}, \bibinfo
  {author} {\bibfnamefont {Y.}~\bibnamefont {Ji}}, \ and\ \bibinfo {author}
  {\bibfnamefont {Y.-M.}\ \bibnamefont {Wang}},\ }\href {\doibase
  10.1103/PhysRevLett.128.062003} {\bibfield  {journal} {\bibinfo  {journal}
  {Phys. Rev. Lett.}\ }\textbf {\bibinfo {volume} {128}},\ \bibinfo {pages}
  {062003} (\bibinfo {year} {2022})},\ \Eprint
  {http://arxiv.org/abs/2106.01390} {arXiv:2106.01390 [hep-ph]} \BibitemShut
  {NoStop}%
\bibitem [{\citenamefont {Freund}\ and\ \citenamefont
  {McDermott}(2002)}]{Freund:2001rk}%
  \BibitemOpen
  \bibfield  {author} {\bibinfo {author} {\bibfnamefont {A.}~\bibnamefont
  {Freund}}\ and\ \bibinfo {author} {\bibfnamefont {M.~F.}\ \bibnamefont
  {McDermott}},\ }\href {\doibase 10.1103/PhysRevD.65.074008} {\bibfield
  {journal} {\bibinfo  {journal} {Phys. Rev. D}\ }\textbf {\bibinfo {volume}
  {65}},\ \bibinfo {pages} {074008} (\bibinfo {year} {2002})},\ \Eprint
  {http://arxiv.org/abs/hep-ph/0106319} {arXiv:hep-ph/0106319} \BibitemShut
  {NoStop}%
\bibitem [{\citenamefont {Moutarde}\ \emph {et~al.}(2013)\citenamefont
  {Moutarde}, \citenamefont {Pire}, \citenamefont {Sabatie}, \citenamefont
  {Szymanowski},\ and\ \citenamefont {Wagner}}]{Moutarde:2013qs}%
  \BibitemOpen
  \bibfield  {author} {\bibinfo {author} {\bibfnamefont {H.}~\bibnamefont
  {Moutarde}}, \bibinfo {author} {\bibfnamefont {B.}~\bibnamefont {Pire}},
  \bibinfo {author} {\bibfnamefont {F.}~\bibnamefont {Sabatie}}, \bibinfo
  {author} {\bibfnamefont {L.}~\bibnamefont {Szymanowski}}, \ and\ \bibinfo
  {author} {\bibfnamefont {J.}~\bibnamefont {Wagner}},\ }\href {\doibase
  10.1103/PhysRevD.87.054029} {\bibfield  {journal} {\bibinfo  {journal} {Phys.
  Rev. D}\ }\textbf {\bibinfo {volume} {87}},\ \bibinfo {pages} {054029}
  (\bibinfo {year} {2013})},\ \Eprint {http://arxiv.org/abs/1301.3819}
  {arXiv:1301.3819 [hep-ph]} \BibitemShut {NoStop}%
\bibitem [{\citenamefont {Braun}\ \emph {et~al.}(2014)\citenamefont {Braun},
  \citenamefont {Manashov}, \citenamefont {M\"uller},\ and\ \citenamefont
  {Pirnay}}]{Braun:2014sta}%
  \BibitemOpen
  \bibfield  {author} {\bibinfo {author} {\bibfnamefont {V.~M.}\ \bibnamefont
  {Braun}}, \bibinfo {author} {\bibfnamefont {A.~N.}\ \bibnamefont {Manashov}},
  \bibinfo {author} {\bibfnamefont {D.}~\bibnamefont {M\"uller}}, \ and\
  \bibinfo {author} {\bibfnamefont {B.~M.}\ \bibnamefont {Pirnay}},\ }\href
  {\doibase 10.1103/PhysRevD.89.074022} {\bibfield  {journal} {\bibinfo
  {journal} {Phys. Rev. D}\ }\textbf {\bibinfo {volume} {89}},\ \bibinfo
  {pages} {074022} (\bibinfo {year} {2014})},\ \Eprint
  {http://arxiv.org/abs/1401.7621} {arXiv:1401.7621 [hep-ph]} \BibitemShut
  {NoStop}%
\bibitem [{\citenamefont {Guo}\ \emph {et~al.}(2021)\citenamefont {Guo},
  \citenamefont {Ji},\ and\ \citenamefont {Shiells}}]{Guo:2021gru}%
  \BibitemOpen
  \bibfield  {author} {\bibinfo {author} {\bibfnamefont {Y.}~\bibnamefont
  {Guo}}, \bibinfo {author} {\bibfnamefont {X.}~\bibnamefont {Ji}}, \ and\
  \bibinfo {author} {\bibfnamefont {K.}~\bibnamefont {Shiells}},\ }\href
  {\doibase 10.1007/JHEP12(2021)103} {\bibfield  {journal} {\bibinfo  {journal}
  {JHEP}\ }\textbf {\bibinfo {volume} {12}},\ \bibinfo {pages} {103} (\bibinfo
  {year} {2021})},\ \Eprint {http://arxiv.org/abs/2109.10373} {arXiv:2109.10373
  [hep-ph]} \BibitemShut {NoStop}%
\bibitem [{\citenamefont {Diehl}(2003)}]{Diehl:2003ny}%
  \BibitemOpen
  \bibfield  {author} {\bibinfo {author} {\bibfnamefont {M.}~\bibnamefont
  {Diehl}},\ }\href {\doibase 10.1016/j.physrep.2003.08.002} {\bibfield
  {journal} {\bibinfo  {journal} {Phys. Rept.}\ }\textbf {\bibinfo {volume}
  {388}},\ \bibinfo {pages} {41} (\bibinfo {year} {2003})},\ \Eprint
  {http://arxiv.org/abs/hep-ph/0307382} {arXiv:hep-ph/0307382} \BibitemShut
  {NoStop}%
\bibitem [{\citenamefont {Belitsky}\ and\ \citenamefont
  {Radyushkin}(2005)}]{Belitsky:2005qn}%
  \BibitemOpen
  \bibfield  {author} {\bibinfo {author} {\bibfnamefont {A.~V.}\ \bibnamefont
  {Belitsky}}\ and\ \bibinfo {author} {\bibfnamefont {A.~V.}\ \bibnamefont
  {Radyushkin}},\ }\href {\doibase 10.1016/j.physrep.2005.06.002} {\bibfield
  {journal} {\bibinfo  {journal} {Phys. Rept.}\ }\textbf {\bibinfo {volume}
  {418}},\ \bibinfo {pages} {1} (\bibinfo {year} {2005})},\ \Eprint
  {http://arxiv.org/abs/hep-ph/0504030} {arXiv:hep-ph/0504030} \BibitemShut
  {NoStop}%
\bibitem [{\citenamefont {Nogueira}(1993)}]{Nogueira:1991ex}%
  \BibitemOpen
  \bibfield  {author} {\bibinfo {author} {\bibfnamefont {P.}~\bibnamefont
  {Nogueira}},\ }\href {\doibase 10.1006/jcph.1993.1074} {\bibfield  {journal}
  {\bibinfo  {journal} {J. Comput. Phys.}\ }\textbf {\bibinfo {volume} {105}},\
  \bibinfo {pages} {279} (\bibinfo {year} {1993})}\BibitemShut {NoStop}%
\bibitem [{\citenamefont {Vermaseren}(2000)}]{Vermaseren:2000nd}%
  \BibitemOpen
  \bibfield  {author} {\bibinfo {author} {\bibfnamefont {J.~A.~M.}\
  \bibnamefont {Vermaseren}},\ }\href@noop {} {\  (\bibinfo {year} {2000})},\
  \Eprint {http://arxiv.org/abs/math-ph/0010025} {arXiv:math-ph/0010025}
  \BibitemShut {NoStop}%
\bibitem [{\citenamefont {Smirnov}\ and\ \citenamefont
  {Chuharev}(2020)}]{Smirnov:2019qkx}%
  \BibitemOpen
  \bibfield  {author} {\bibinfo {author} {\bibfnamefont {A.~V.}\ \bibnamefont
  {Smirnov}}\ and\ \bibinfo {author} {\bibfnamefont {F.~S.}\ \bibnamefont
  {Chuharev}},\ }\href {\doibase 10.1016/j.cpc.2019.106877} {\bibfield
  {journal} {\bibinfo  {journal} {Comput. Phys. Commun.}\ }\textbf {\bibinfo
  {volume} {247}},\ \bibinfo {pages} {106877} (\bibinfo {year} {2020})},\
  \Eprint {http://arxiv.org/abs/1901.07808} {arXiv:1901.07808 [hep-ph]}
  \BibitemShut {NoStop}%
\bibitem [{\citenamefont {Gao}\ \emph {et~al.}(2021)\citenamefont {Gao},
  \citenamefont {Huber}, \citenamefont {Ji},\ and\ \citenamefont
  {Wang}}]{Gao:2021beo}%
  \BibitemOpen
  \bibfield  {author} {\bibinfo {author} {\bibfnamefont {J.}~\bibnamefont
  {Gao}}, \bibinfo {author} {\bibfnamefont {T.}~\bibnamefont {Huber}}, \bibinfo
  {author} {\bibfnamefont {Y.}~\bibnamefont {Ji}}, \ and\ \bibinfo {author}
  {\bibfnamefont {Y.-M.}\ \bibnamefont {Wang}},\ }in\ \href@noop {} {\emph
  {\bibinfo {booktitle} {{15th International Symposium on Radiative
  Corrections: Applications of Quantum Field Theory to Phenomenology AND
  LoopFest XIX: Workshop on Radiative Corrections for the LHC and Future
  Colliders}}}}\ (\bibinfo {year} {2021})\ \Eprint
  {http://arxiv.org/abs/2110.14776} {arXiv:2110.14776 [hep-ph]} \BibitemShut
  {NoStop}%
\bibitem [{\citenamefont {Panzer}(2015)}]{Panzer:2014caa}%
  \BibitemOpen
  \bibfield  {author} {\bibinfo {author} {\bibfnamefont {E.}~\bibnamefont
  {Panzer}},\ }\href {\doibase 10.1016/j.cpc.2014.10.019} {\bibfield  {journal}
  {\bibinfo  {journal} {Comput. Phys. Commun.}\ }\textbf {\bibinfo {volume}
  {188}},\ \bibinfo {pages} {148} (\bibinfo {year} {2015})},\ \Eprint
  {http://arxiv.org/abs/1403.3385} {arXiv:1403.3385 [hep-th]} \BibitemShut
  {NoStop}%
\bibitem [{\citenamefont {Maitre}(2006)}]{Maitre:2005uu}%
  \BibitemOpen
  \bibfield  {author} {\bibinfo {author} {\bibfnamefont {D.}~\bibnamefont
  {Maitre}},\ }\href {\doibase 10.1016/j.cpc.2005.10.008} {\bibfield  {journal}
  {\bibinfo  {journal} {Comput. Phys. Commun.}\ }\textbf {\bibinfo {volume}
  {174}},\ \bibinfo {pages} {222} (\bibinfo {year} {2006})},\ \Eprint
  {http://arxiv.org/abs/hep-ph/0507152} {arXiv:hep-ph/0507152} \BibitemShut
  {NoStop}%
\bibitem [{\citenamefont {Remiddi}\ and\ \citenamefont
  {Vermaseren}(2000)}]{Remiddi:1999ew}%
  \BibitemOpen
  \bibfield  {author} {\bibinfo {author} {\bibfnamefont {E.}~\bibnamefont
  {Remiddi}}\ and\ \bibinfo {author} {\bibfnamefont {J.~A.~M.}\ \bibnamefont
  {Vermaseren}},\ }\href {\doibase 10.1142/S0217751X00000367} {\bibfield
  {journal} {\bibinfo  {journal} {Int. J. Mod. Phys. A}\ }\textbf {\bibinfo
  {volume} {15}},\ \bibinfo {pages} {725} (\bibinfo {year} {2000})},\ \Eprint
  {http://arxiv.org/abs/hep-ph/9905237} {arXiv:hep-ph/9905237} \BibitemShut
  {NoStop}%
\bibitem [{\citenamefont {Altinoluk}\ \emph {et~al.}(2012)\citenamefont
  {Altinoluk}, \citenamefont {Pire}, \citenamefont {Szymanowski},\ and\
  \citenamefont {Wallon}}]{Altinoluk:2012nt}%
  \BibitemOpen
  \bibfield  {author} {\bibinfo {author} {\bibfnamefont {T.}~\bibnamefont
  {Altinoluk}}, \bibinfo {author} {\bibfnamefont {B.}~\bibnamefont {Pire}},
  \bibinfo {author} {\bibfnamefont {L.}~\bibnamefont {Szymanowski}}, \ and\
  \bibinfo {author} {\bibfnamefont {S.}~\bibnamefont {Wallon}},\ }\href
  {\doibase 10.1007/JHEP10(2012)049} {\bibfield  {journal} {\bibinfo  {journal}
  {JHEP}\ }\textbf {\bibinfo {volume} {10}},\ \bibinfo {pages} {049} (\bibinfo
  {year} {2012})},\ \Eprint {http://arxiv.org/abs/1207.4609} {arXiv:1207.4609
  [hep-ph]} \BibitemShut {NoStop}%
\bibitem [{\citenamefont {Radyushkin}(1999)}]{Radyushkin:1998bz}%
  \BibitemOpen
  \bibfield  {author} {\bibinfo {author} {\bibfnamefont {A.~V.}\ \bibnamefont
  {Radyushkin}},\ }\href {\doibase 10.1016/S0370-2693(98)01584-6} {\bibfield
  {journal} {\bibinfo  {journal} {Phys. Lett. B}\ }\textbf {\bibinfo {volume}
  {449}},\ \bibinfo {pages} {81} (\bibinfo {year} {1999})},\ \Eprint
  {http://arxiv.org/abs/hep-ph/9810466} {arXiv:hep-ph/9810466} \BibitemShut
  {NoStop}%
\bibitem [{\citenamefont {Polyakov}\ and\ \citenamefont
  {Weiss}(1999)}]{Polyakov:1999gs}%
  \BibitemOpen
  \bibfield  {author} {\bibinfo {author} {\bibfnamefont {M.~V.}\ \bibnamefont
  {Polyakov}}\ and\ \bibinfo {author} {\bibfnamefont {C.}~\bibnamefont
  {Weiss}},\ }\href {\doibase 10.1103/PhysRevD.60.114017} {\bibfield  {journal}
  {\bibinfo  {journal} {Phys. Rev. D}\ }\textbf {\bibinfo {volume} {60}},\
  \bibinfo {pages} {114017} (\bibinfo {year} {1999})},\ \Eprint
  {http://arxiv.org/abs/hep-ph/9902451} {arXiv:hep-ph/9902451} \BibitemShut
  {NoStop}%
\bibitem [{\citenamefont {Musatov}\ and\ \citenamefont
  {Radyushkin}(2000)}]{Musatov:1999xp}%
  \BibitemOpen
  \bibfield  {author} {\bibinfo {author} {\bibfnamefont {I.~V.}\ \bibnamefont
  {Musatov}}\ and\ \bibinfo {author} {\bibfnamefont {A.~V.}\ \bibnamefont
  {Radyushkin}},\ }\href {\doibase 10.1103/PhysRevD.61.074027} {\bibfield
  {journal} {\bibinfo  {journal} {Phys. Rev. D}\ }\textbf {\bibinfo {volume}
  {61}},\ \bibinfo {pages} {074027} (\bibinfo {year} {2000})},\ \Eprint
  {http://arxiv.org/abs/hep-ph/9905376} {arXiv:hep-ph/9905376} \BibitemShut
  {NoStop}%
\bibitem [{\citenamefont {Goloskokov}\ and\ \citenamefont
  {Kroll}(2007)}]{Goloskokov:2006hr}%
  \BibitemOpen
  \bibfield  {author} {\bibinfo {author} {\bibfnamefont {S.~V.}\ \bibnamefont
  {Goloskokov}}\ and\ \bibinfo {author} {\bibfnamefont {P.}~\bibnamefont
  {Kroll}},\ }\href {\doibase 10.1140/epjc/s10052-007-0228-4} {\bibfield
  {journal} {\bibinfo  {journal} {Eur. Phys. J. C}\ }\textbf {\bibinfo {volume}
  {50}},\ \bibinfo {pages} {829} (\bibinfo {year} {2007})},\ \Eprint
  {http://arxiv.org/abs/hep-ph/0611290} {arXiv:hep-ph/0611290} \BibitemShut
  {NoStop}%
\bibitem [{\citenamefont {Kroll}\ \emph {et~al.}(2013)\citenamefont {Kroll},
  \citenamefont {Moutarde},\ and\ \citenamefont {Sabatie}}]{Kroll:2012sm}%
  \BibitemOpen
  \bibfield  {author} {\bibinfo {author} {\bibfnamefont {P.}~\bibnamefont
  {Kroll}}, \bibinfo {author} {\bibfnamefont {H.}~\bibnamefont {Moutarde}}, \
  and\ \bibinfo {author} {\bibfnamefont {F.}~\bibnamefont {Sabatie}},\ }\href
  {\doibase 10.1140/epjc/s10052-013-2278-0} {\bibfield  {journal} {\bibinfo
  {journal} {Eur. Phys. J. C}\ }\textbf {\bibinfo {volume} {73}},\ \bibinfo
  {pages} {2278} (\bibinfo {year} {2013})},\ \Eprint
  {http://arxiv.org/abs/1210.6975} {arXiv:1210.6975 [hep-ph]} \BibitemShut
  {NoStop}%
\bibitem [{\citenamefont {Abramowicz}\ \emph {et~al.}(2015)\citenamefont
  {Abramowicz} \emph {et~al.}}]{H1:2015ubc}%
  \BibitemOpen
  \bibfield  {author} {\bibinfo {author} {\bibfnamefont {H.}~\bibnamefont
  {Abramowicz}} \emph {et~al.} (\bibinfo {collaboration} {H1, ZEUS}),\ }\href
  {\doibase 10.1140/epjc/s10052-015-3710-4} {\bibfield  {journal} {\bibinfo
  {journal} {Eur. Phys. J. C}\ }\textbf {\bibinfo {volume} {75}},\ \bibinfo
  {pages} {580} (\bibinfo {year} {2015})},\ \Eprint
  {http://arxiv.org/abs/1506.06042} {arXiv:1506.06042 [hep-ex]} \BibitemShut
  {NoStop}%
\bibitem [{\citenamefont {Alekhin}\ \emph {et~al.}(2017)\citenamefont
  {Alekhin}, \citenamefont {Bl\"umlein}, \citenamefont {Moch},\ and\
  \citenamefont {Placakyte}}]{Alekhin:2017kpj}%
  \BibitemOpen
  \bibfield  {author} {\bibinfo {author} {\bibfnamefont {S.}~\bibnamefont
  {Alekhin}}, \bibinfo {author} {\bibfnamefont {J.}~\bibnamefont {Bl\"umlein}},
  \bibinfo {author} {\bibfnamefont {S.}~\bibnamefont {Moch}}, \ and\ \bibinfo
  {author} {\bibfnamefont {R.}~\bibnamefont {Placakyte}},\ }\href {\doibase
  10.1103/PhysRevD.96.014011} {\bibfield  {journal} {\bibinfo  {journal} {Phys.
  Rev. D}\ }\textbf {\bibinfo {volume} {96}},\ \bibinfo {pages} {014011}
  (\bibinfo {year} {2017})},\ \Eprint {http://arxiv.org/abs/1701.05838}
  {arXiv:1701.05838 [hep-ph]} \BibitemShut {NoStop}%
\bibitem [{\citenamefont {Alekhin}\ \emph {et~al.}(2018)\citenamefont
  {Alekhin}, \citenamefont {Bl\"umlein},\ and\ \citenamefont
  {Moch}}]{Alekhin:2018pai}%
  \BibitemOpen
  \bibfield  {author} {\bibinfo {author} {\bibfnamefont {S.}~\bibnamefont
  {Alekhin}}, \bibinfo {author} {\bibfnamefont {J.}~\bibnamefont {Bl\"umlein}},
  \ and\ \bibinfo {author} {\bibfnamefont {S.}~\bibnamefont {Moch}},\ }\href
  {\doibase 10.1140/epjc/s10052-018-5947-1} {\bibfield  {journal} {\bibinfo
  {journal} {Eur. Phys. J. C}\ }\textbf {\bibinfo {volume} {78}},\ \bibinfo
  {pages} {477} (\bibinfo {year} {2018})},\ \Eprint
  {http://arxiv.org/abs/1803.07537} {arXiv:1803.07537 [hep-ph]} \BibitemShut
  {NoStop}%
\bibitem [{\citenamefont {Vinnikov}(2006)}]{Vinnikov:2006xw}%
  \BibitemOpen
  \bibfield  {author} {\bibinfo {author} {\bibfnamefont {A.~V.}\ \bibnamefont
  {Vinnikov}},\ }\href@noop {} {\  (\bibinfo {year} {2006})},\ \Eprint
  {http://arxiv.org/abs/hep-ph/0604248} {arXiv:hep-ph/0604248} \BibitemShut
  {NoStop}%
\bibitem [{\citenamefont {Bertone}\ \emph {et~al.}(2022)\citenamefont
  {Bertone}, \citenamefont {Dutrieux}, \citenamefont {Mezrag}, \citenamefont
  {Morgado},\ and\ \citenamefont {Moutarde}}]{Bertone:2022frx}%
  \BibitemOpen
  \bibfield  {author} {\bibinfo {author} {\bibfnamefont {V.}~\bibnamefont
  {Bertone}}, \bibinfo {author} {\bibfnamefont {H.}~\bibnamefont {Dutrieux}},
  \bibinfo {author} {\bibfnamefont {C.}~\bibnamefont {Mezrag}}, \bibinfo
  {author} {\bibfnamefont {J.~M.}\ \bibnamefont {Morgado}}, \ and\ \bibinfo
  {author} {\bibfnamefont {H.}~\bibnamefont {Moutarde}},\ }\href@noop {} {\
  (\bibinfo {year} {2022})},\ \Eprint {http://arxiv.org/abs/2206.01412}
  {arXiv:2206.01412 [hep-ph]} \BibitemShut {NoStop}%
\bibitem [{\citenamefont {Berthou}\ \emph {et~al.}(2018)\citenamefont {Berthou}
  \emph {et~al.}}]{Berthou:2015oaw}%
  \BibitemOpen
  \bibfield  {author} {\bibinfo {author} {\bibfnamefont {B.}~\bibnamefont
  {Berthou}} \emph {et~al.},\ }\href {\doibase 10.1140/epjc/s10052-018-5948-0}
  {\bibfield  {journal} {\bibinfo  {journal} {Eur. Phys. J. C}\ }\textbf
  {\bibinfo {volume} {78}},\ \bibinfo {pages} {478} (\bibinfo {year} {2018})},\
  \Eprint {http://arxiv.org/abs/1512.06174} {arXiv:1512.06174 [hep-ph]}
  \BibitemShut {NoStop}%
\end{thebibliography}

%

\pagebreak
\widetext
\begin{center}
\textbf{\large Supplemental Materials}
\end{center}
\setcounter{equation}{0}
\setcounter{figure}{0}
\setcounter{table}{0}
\setcounter{page}{1}
\makeatletter
\renewcommand{\theequation}{S\arabic{equation}}
\renewcommand{\thefigure}{S\arabic{figure}}

%
\appendix

\begin{widetext}
\section{DVCS coefficient functions in the \texorpdfstring{$\widebar{\rm MS}$}{Lg}-scheme }
%
\label{app:CFs}
In this Appendix we collect explicit expressions for the CFs. The notations correspond to 
Eqs. (\ref{eq:CF1}), (\ref{eq: CF2quark}), (\ref{eq:CF2}).
In what follows
\begin{align}
    z = \frac{1}{2} (1-x/\xi)\,, && L = \ln \frac{\mu^2}{Q^2}\,.
\end{align}  
For completeness, we present also the  well-known tree-level and one-loop expressions: 
\begin{align}
C_q^{(0)} &= \frac{e_q^2 (1-2z)}{2z(1-z)} ,
\\
C_q^{(1)} &=  \frac{ e_q^2 C_F}{2z (1-z) } \biggl \{ L \Big [ 4(z \ln (1-z) - (1-z) \ln z) - 3 (1-2z)\Big ] +   (1-z) \ln^2 z - z \ln^2 (1-z) 
\notag\\
&\quad + 3 \Big [(1-z) \ln (1-z) -  z \ln z \Big ]  - 9 (1-2z) \biggr \} 
\\
C_g^{(1)} &=  \frac{\Big ( \sum_q e_q^2 \Big ) T_F}{4z^2 (1-z)^2}  \biggl \{ 2L \Big [ z^2 \ln z + (1-z)^2 \ln (1-z) \Big ] - z^2 \ln^2 z - (1-z)^2 \ln^2 (1-z) 
\notag\\
&\quad+ 2 \Big [ z(1+z) \ln z + (1-z) (2-z) \ln (1-z) \Big ] \biggr \}.
\end{align}
The two-loop CFs are written in terms of harmonic polylogarithms $ H_{i,j,\ldots} \equiv H_{i,j,\ldots}(z)$ \cite{Remiddi:1999ew}
{\allowdisplaybreaks
\begin{align}
\text C_{\text{NS}}^{(F)} &=  L^2 \Big[ -4 z H_{1,1}-4 (z-1) H_{0,0}-2 (z-1) H_{1,0}-2 H_2 z+H_0 (6-4 z)+H_1 (4
   z+2)+2 \zeta _2 z-9 z+\frac{9}{2} \Big]
\notag\\
&\quad + L \Big[ -8 z H_{1,2}-12 z H_{1,1,1}+(8 z-6) H_{0,0}-8 (z-1) H_{1,0}+(8 z-2)
   H_{1,1}+8 (z-1) H_{2,0}+12 (z-1) H_{0,0,0}
\notag\\
&\quad +4 (z+1) H_{1,1,0}+H_1 \Big(4
   \zeta _2 (3 z+1)+7 z+11\Big)-8 H_2 z+H_0 (18-7 z)-4 H_3 (z-2)+\zeta _2 (6-4 z)
\notag\\
&\quad +28 \zeta _3 (z-1) -51 z+\frac{51}{2} \Big]
\notag\\
&\quad -12 (z-1) H_{0,0,0,0}+(36-58 z) H_{1,0,0,0}+(40-70 z) H_{1,1,0,0}-6 (7 z-6)
   H_{1,1,1,0}-12 z H_{1,1,1,1}
\notag\\
&\quad -24 (z-1) \Big(2 z^2-z+1\Big) H_{1,1,0}-2 z H_{0,0,0}+4 (3 z-2)
   H_{1,0,0}+2 (z-1) H_{1,1,1}+(42-72 z) H_{1,1,2}
\notag\\
&\quad +(28-38 z)
   H_{1,2,0}+(38-68 z) H_{1,2,1}+(46-68 z) H_{2,0,0}+(44-70 z)
   H_{2,1,0}+(38-58 z) H_{2,1,1}
\notag\\
&\quad + H_{1,0} \Big(48 z^2+\zeta _2 (28-38 z)-44 z+10\Big)+6 (z-1) \Big(8 z^2-4
   z+3\Big) H_{1,2}-6 z \Big(8 z^2-12 z+7\Big) H_{2,0}
\notag\\
&\quad +H_{1,1}
   \Big(\zeta _2 (26 z-2)+25 z-7\Big)+H_{0,0} \Big((60 z-38) H_{1,1}-4
   \zeta _2 z+25 z-18\Big)+(42-70 z) H_{1,3}
\notag\\
&\quad +(4-12 z) H_{2,1}+(26-38 z)
   H_{2,2}+(46-72 z) H_{3,0}+(46-70 z) H_{3,1} 
\notag\\
&\quad + H_2 \Big(-48 z^2+\zeta _2 (44-70 z)+52 z-14\Big) +24 H_3 z \Big(2 z^2-3 z+2\Big)+H_4 (22-42 z)
\notag\\
&\quad +H_0 \Big(\zeta _2
   \Big(H_1 (60 z-38)-48 z^3+72 z^2-44 z+6\Big)+\zeta _3 (8
   z+4)+\frac{1}{2} (121-96 z) z\Big)
\notag\\
&\quad +H_1 \Big(\zeta _2 \Big(-48 z^3+72
   z^2-46 z+28\Big)+\frac{1}{2} \Big(-96 z^2+\zeta _3 (44 z-8)+71
   z+25\Big)\Big)
\notag\\
&\quad + \zeta _2 \Big(48 z^2-76 z+26\Big)+\zeta _3 \Big(144 z^3-216 z^2+154
   z-69\Big)-\frac{1}{5} \zeta _2^2 (31 z-13)-\frac{331}{8} (2 z-1),
\end{align}
%
%
\begin{align}
\text C_{\text{NS}}^{(A)} &= L \Big[4 z H_{1,2}-4 z H_{1,1,0}-4 (z-1) H_{2,0}-\frac{4}{3} H_0 \Big(3 \zeta _2
   (z-1)-5 z+2\Big)-\frac{4}{3} H_1 \Big(3 \zeta _2 z+5 z-3\Big)+4 H_3
   (z-1)
\notag\\
&\quad -12 \zeta _3 (z-1)+2 z-1 \Big]
\notag\\
&\quad + (30 z-19) H_{1,0,0,0}+(36 z-21) H_{1,1,0,0}+(20 z-17) H_{1,1,1,0}
\notag\\
&\quad + 2 (z-1) \Big(12 z^2-6 z+5\Big) H_{1,1,0}+(36 z-21) H_{1,1,2}+5 (4 z-3)
   H_{1,2,0}+(36 z-19) H_{1,2,1}
\notag\\
&\quad +(36 z-25) H_{2,0,0}+(36 z-23) H_{2,1,0}+(30
   z-19) H_{2,1,1}
\notag\\
&\quad + H_{1,0} \Big(-24 z^2+5 \zeta _2 (4 z-3)+\frac{64 z}{3}-\frac{4}{3}\Big)-2
   (z-1) \Big(12 z^2-6 z+5\Big) H_{1,2}+2 z \Big(12 z^2-18 z+11\Big)
   H_{2,0}
\notag\\
&\quad +H_{1,1} \Big(\zeta _2 (2-10 z)-\frac{38 z}{3}+10\Big)+H_{0,0}
   \Big((19-30 z) H_{1,1}+\frac{2}{3} \Big(\zeta _2 (9 z-6)-19
   z+4\Big)\Big)+(36 z-21) H_{1,3}
\notag\\
&\quad +(20 z-13) H_{2,2}+(36 z-23)
   H_{3,0}+(36 z-23) H_{3,1}
\notag\\
&\quad + H_2 \Big(24 z^2+\zeta _2 (36 z-23)-\frac{80 z}{3}+4\Big)+H_0 \Big(2
   \zeta _2 \Big(12 z^2-18 z+11\Big) z+24 z^2+\zeta _3 (14-20 z)-\frac{73
   z}{9}-\frac{32}{9}\Big)
\notag\\
&\quad +H_1 \Big(-\zeta _2 H_0 (30 z-19)+2 \zeta _2
   \Big(12 z^3-18 z^2+11 z-5\Big)+24 z^2+4 \zeta _3 z-\frac{359
   z}{9}+\frac{37}{3}\Big)
\notag\\
&\quad -2 H_3 z \Big(12 z^2-18 z+11\Big)+H_4 (20
   z-11)
\notag\\
&\quad -\frac{2}{3} \zeta _2 \Big(36 z^2-50 z+11\Big)-6 \zeta _3 \Big(12 z^3-18
   z^2+19 z-9\Big)+\frac{1}{5} \zeta _2^2 (13 z+1)+\frac{73}{12} (2 z-1),
\end{align}
%
%
\begin{align}
\text C_{\text{NS}}^{(\beta_0)} &= L^2 \Big[-H_1 z+H_0 (z-1)+3 z-\frac{3}{2} \Big]
\notag\\
&\quad + L \Big[-2 z H_{1,1}-2 (z-1) H_{0,0}+2 (z-1) H_{1,0}+2 H_2 z+\frac{1}{3} H_0 (7
   z-10)-\frac{1}{3} H_1 (7 z+3)+2 \zeta _2 (z-1)+19 z-\frac{19}{2} \Big]
\notag\\
&\quad + 2 (z-1) H_{0,0,0}+(2-2 z) H_{1,0,0}+(z-1) H_{1,1,0}-2 z H_{1,1,1}
\notag\\
&\quad + \frac{1}{3} (10-7 z) H_{0,0}+\frac{2}{3} (4 z-7) H_{1,0}+\Big(-\frac{7
   z}{3}-1\Big) H_{1,1}+2 z H_{2,1}
\notag\\
&\quad + H_1 \Big(\zeta _2 (z-1)+\frac{1}{18} (31 z-69)\Big)+\frac{1}{18} H_0 (-31
   z-38)+H_2 \Big(\frac{8 z}{3}+2\Big)-H_3 z
\notag\\
&\quad + \frac{2}{3} \zeta _2 (4 z-7)+\zeta _3 (z-1)+\frac{457}{24} (2 z-1),
\end{align}
These expressions agree identically with the corresponding results in Ref.~\cite{Braun:2020yib}.
The expressions below for the two-loop pure-singlet quark and gluon CFs are new results. 
%
%
\begin{align}
\text C_{\mathrm PS} &= L^2  \Big[ -8 (z-1) H_{1,0}+4 H_1 \Big(4 z^2-5 z+1\Big)+4 H_0 z (4 z-3)-8 H_2 z+8
   \zeta _2 z \Big]
\notag\\
&\quad + 8L  \Big[-z (4 z-3) H_{0,0}-(z-1) (4 z-1) H_{1,0}+(z-1) (4 z-1) H_{1,1}-2 z
   H_{2,1}+2 (z-1) H_{1,0,0}-2 (z-1) H_{1,1,0}
\notag\\
&\quad -\Big(2 \zeta _2-3\Big) H_1
   (z-1)-3 H_0 z+H_2 z (4 z-3)+2 H_3 z-\zeta _2 z (4 z-3)\Big]
\notag\\
&\quad -16 (z-1) H_{1,0,0,0}+16 (z-1) H_{1,1,0,0}+(8-8 z) H_{1,1,1,0}
\notag\\[1mm]
&\quad + 8 z (4 z-3) H_{0,0,0}+8 (z-1) (4 z-1) H_{1,0,0}-4 (z-1) (4 z+5) H_{1,1,0}+8
   (z-1) (4 z-1) H_{1,1,1}-16 z H_{2,1,1}
\notag\\
&\quad -16 z^2 H_{2,0}-8 \Big(\zeta _2-3\Big) (z-1) H_{1,1}-16 (z-1)^2
   H_{1,2}+24 z H_{0,0}+(4-4 z) H_{1,0}+8 z (4 z-3) H_{2,1}+16 z H_{3,1}
\notag\\
&\quad -8 H_0 z \Big(2 \zeta _2 z+5\Big)-4 H_1 (z-1) \Big(2 \Big(\zeta
   _3-5\Big)+\zeta _2 (4 z+5)\Big)-4 H_2 z-4 H_3 z (4 z-9)-8 H_4 z
\notag\\
&\quad + 4 z \Big(2 \zeta _2^2+\zeta _2-3 \zeta _3 (4 z+1)\Big),
\end{align}


%
%
\begin{align}
\text C_{g}^{(F)} &= L^2 \Big[ -2 \Big(z^2 H_{0,0}+(z-1) \Big((z-1) H_{1,1}+z\Big)\Big)+H_1
   \Big(z^2-1\Big)+H_0 (-(z-2)) z \Big]
\notag\\
&\quad + L \Big[ 8 z^2 H_{0,0,0}-4 (z-1)^2 H_{1,0}-8 (z-1)^2 H_{1,1,1}-4 (z-3) (z-1)
   H_{1,1}-4 z (z+2) H_{0,0}-4 H_2 z^2
\notag\\
&\quad +4 H_1 (z-1) \Big(\zeta _2 (z-1)-2
   z+5\Big)-4 H_0 z \Big(\zeta _2 z-2 z-3\Big)+4 z \Big(\zeta _2 z-4
   z+4\Big) \Big]
\notag\\
&\quad -10 z^2 H_{0,0,0,0}-36 z^2 H_{1,0,0,0}-36 z^2 H_{1,1,0,0}-4 \Big(7 z^2+4
   z-2\Big) H_{1,1,1,0}-10 (z-1)^2 H_{1,1,1,1}
\notag\\
&\quad + \Big(-8 z^4+16 z^3-5 z^2+2 z-5\Big) H_{1,1,0}-2 \Big(23 z^2-10
   z+5\Big) H_{1,1,2}-4 \Big(7 z^2+4 z-2\Big) H_{1,2,0}
\notag\\
&\quad +\Big(-34 z^2-4
   z+2\Big) H_{1,2,1}-34 z^2 H_{2,0,0}-44 z^2 H_{2,1,0}-36 z^2 H_{2,1,1}+4
   (z-1)^2 H_{1,0,0}
\notag\\
&\quad -10 (z-2) (z-1) H_{1,1,1}+10 z (z+1) H_{0,0,0}
\notag\\
&\quad + H_{1,0} \Big(-4 \zeta _2 \Big(7 z^2+4 z-2\Big)+8 z^3-11 z^2+6
   z-3\Big)-4 z^2 H_{2,1}-28 z^2 H_{2,2}-46 z^2 H_{3,0}-36 z^2 H_{3,1}
\notag\\
&\quad +8
   \Big(z^4-2 z^3+z\Big) H_{1,2}-4 \Big(11 z^2-4 z+2\Big) H_{1,3}-8
   \Big(z^4-2 z^3+z\Big) H_{2,0}
\notag\\
&\quad -2 z H_{0,0} \Big(-18 z H_{1,1}+3 \zeta
   _2 z+3 z+10\Big)+2 (z-1) H_{1,1} \Big(6 \zeta _2 (z-1)-3 z+13\Big)
\notag\\
&\quad -H_2 z \Big(8 z^2+44 \zeta _2 z-13 z+8\Big)+2 H_0 z \Big(\zeta _2
   \Big(18 H_1 z-4 z^3+8 z^2-3 z-5\Big)-4 z^2-4 \zeta _3 z+23
   z+9\Big)
\notag\\
&\quad -H_1 (z-1) \Big(\zeta _2 \Big(8 z^3-8 z^2-9 z+3\Big)+8
   z^2-2 \zeta _3 (z-1)+30 z-56\Big)-28 H_4 z^2+H_3 \Big(8 z^3-16 z^2+5
   z+8\Big) z
\notag\\
&\quad + z \Big(\zeta _2 \Big(8 z^2-13 z+8\Big)+\zeta _3 \Big(24 z^3-48
   z^2-z+24\Big)-16 \zeta _2^2 z-36 (z-1)\Big),
\end{align}
%
%
\begin{align}
\text C_{g}^{(A)} &= 2L^2 \Big[ -z^2 H_{0,0}-3 (z-1)^2 H_{1,0}-(z-1)^2 H_{1,1}+H_0 z \Big(4 z^2-5
   z+4\Big)+H_1 \Big(4 z^3-7 z^2+6 z-3\Big)
\notag\\
&\quad -3 H_2 z^2+z \Big(3 \zeta
   _2 z+z-1\Big)  \Big]
\notag\\
&\quad + L \Big[ 16 \Big(z^2-z+1\Big) (z-1) H_{1,1}-16 z \Big(z^2-z+1\Big) H_{0,0}+4
   z^2 H_{2,0}+\Big(-8 z^2-8 z+4\Big) H_{2,1}+4 z^2 H_{0,0,0}
\notag\\
&\quad +4 \Big(2
   z^2-6 z+3\Big) H_{1,0,0}-8 (2 z+1) (z-1)^2 H_{1,0}-4 (z-1)^2 H_{1,2}-8
   (z-1)^2 H_{1,1,0}-4 (z-1)^2 H_{1,1,1}
\notag\\
&\quad +8 H_2 z^2 (2 z-3)+8 H_3 z^2-2 H_1
   (z-1) \Big(2 \zeta _2 (z-1)-7 (z+1)\Big)-14 H_0 (z-2) z
\notag\\
&\quad -2 z \Big(4
   \zeta _2 (2 z-3) z-8 \zeta _3 z-9 z+9\Big) \Big]
\notag\\
&\quad -2 z^2 H_{0,0,0,0}+2 \Big(5 z^2-6 z+3\Big) H_{1,1,0,0}+2 \Big(z^2+6
   z-3\Big) H_{1,1,1,0}-2 (z-1)^2 H_{1,1,1,1}+14 (2 z-1) H_{1,0,0,0}
\notag\\
&\quad + 2 z \Big(8 z^2-8 z+7\Big) H_{0,0,0}+2 \Big(8 z^3-14 z^2-z+4\Big)
   H_{1,0,0}-2 (z-1) \Big(4 z^2+5 z-7\Big) H_{1,1,0}
\notag\\
&\quad +2 (z-1) \Big(8
   z^2-8 z+7\Big) H_{1,1,1}+2 \Big(5 z^2-2 z+1\Big) H_{1,1,2}+2 \Big(5
   z^2+2 z-1\Big) H_{1,2,0}+2 \Big(z^2+6 z-3\Big) H_{1,2,1}
\notag\\
&\quad +2
   \Big(z^2+2 z-1\Big) H_{2,0,0}+2 \Big(5 z^2+2 z-1\Big)
   H_{2,1,0}+(4-8 z) H_{2,1,1}
\notag\\
&\quad + H_{1,0} \Big(2 \zeta _2 \Big(5 z^2+2 z-1\Big)-5 \Big(2 z^2-5
   z+3\Big)\Big)+H_{0,0} \Big(\Big(-6 z^2-4 z+2\Big) H_{1,1}+2 z
   \Big(4 \zeta _2 z+3 z-13\Big)\Big)
\notag\\
&\quad +2 \Big(5 z^2-2 z+1\Big)
   H_{1,3}-4 z^2 (2 z-1) H_{2,0}+2 \Big(8 z^3-10 z^2-5 z+3\Big) H_{2,1}+2
   \Big(5 z^2-2 z+1\Big) H_{2,2}
\notag\\
&\quad +2 \Big(5 z^2+2 z-1\Big) H_{3,0}+2
   \Big(5 z^2+6 z-3\Big) H_{3,1}-4 (2 z-1) (z-1)^2 H_{1,2}+2 (3 z+10)
   (z-1) H_{1,1}
\notag\\
&\quad + H_2 \Big(2 \zeta _2 \Big(5 z^2+2 z-1\Big)-5 z (2 z+1)\Big)
\notag\\
&\quad + H_1\Big(-2 \zeta _2 \Big(4 z^3-z^2-8 z+5\Big)+4 \zeta _3 \Big(5 z^2-4
   z+2\Big)+29 z^2-8 z-21\Big)
\notag\\
&\quad +H_0 \Big(-2 \zeta _2 H_1 \Big(3 z^2+2
   z-1\Big)-z \Big(8 \zeta _2 z^2+4 \zeta _3 z+29 z-50\Big)\Big)-2
   H_3 z \Big(4 z^2-13 z+2\Big)+2 H_4 \Big(z^2+2 z-1\Big)
\notag\\
&\quad + \frac{82}{5} \zeta _2^2 z^2-2 \zeta _3 z \Big(12 z^2-17 z+9\Big)+5 \zeta
   _2 z (2 z+1)+37 (z-1) z.
\end{align}

}  

The $u$-quark $C_u$ and gluon $C_g$ CFs  as defined in Eq.~(\ref{eq:CFF-H})  
at $\mu^2 = Q^2 = 4\text{ GeV}^2$ are shown to the LO/NLO/NNLO accuracy 
in Fig.~\ref{figCF}. For this plot we use $n_f=3$ since, according to the analysis 
in~\cite{Noritzsch:2003un}, the $c$-quark contribution at such scales is very small.   

\begin{figure*}[ht]
~~\includegraphics[scale=0.34]{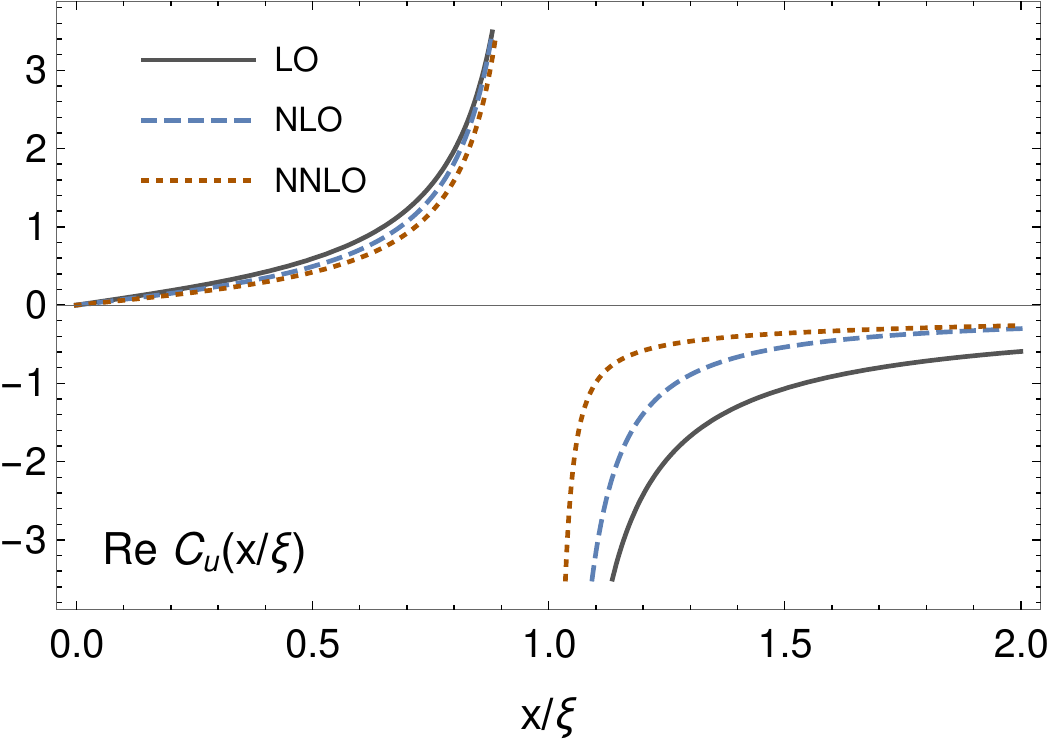} \hspace*{0.4cm}
\includegraphics[scale=0.34]{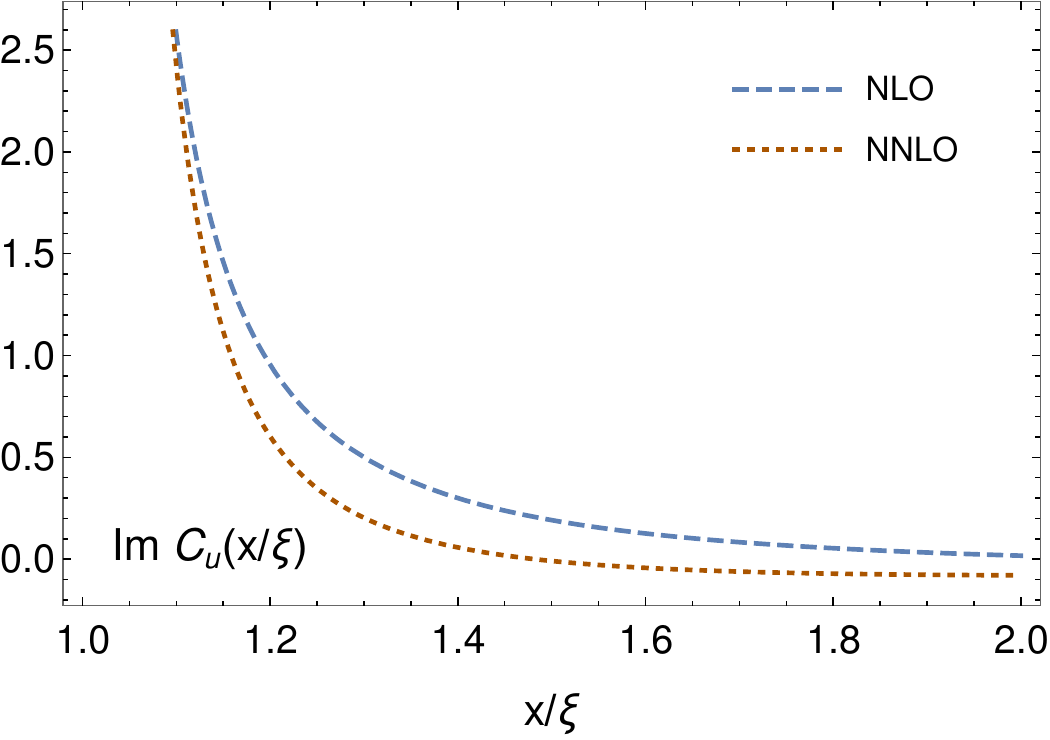} \\
\includegraphics[scale=0.342]{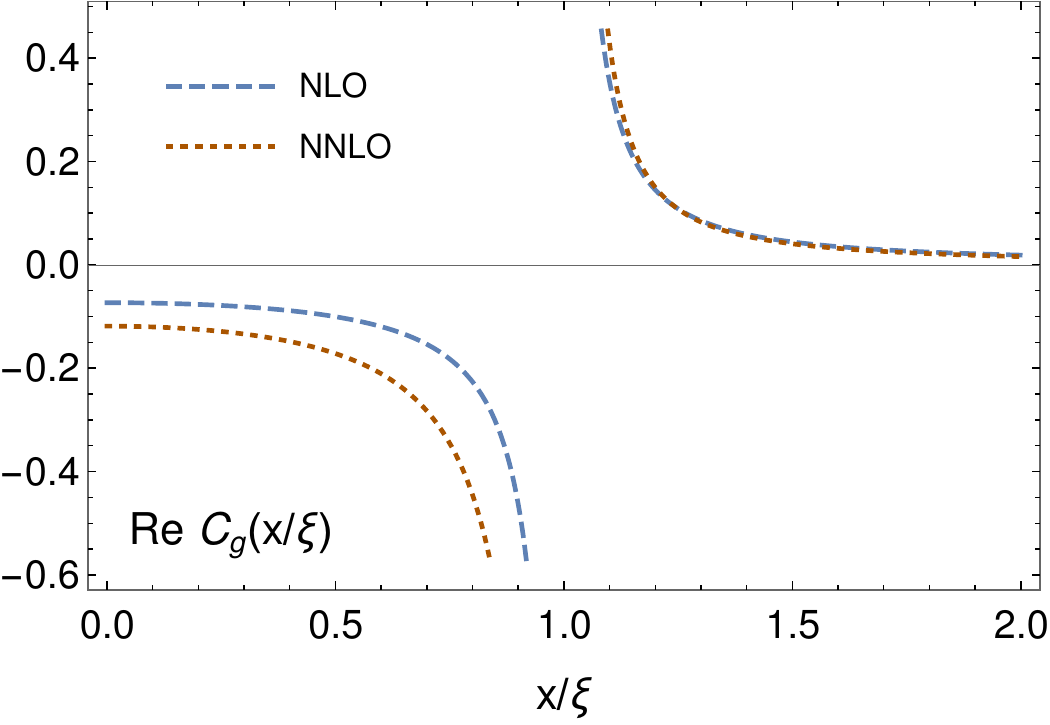} \hspace*{0.4cm}
\includegraphics[scale=0.342]{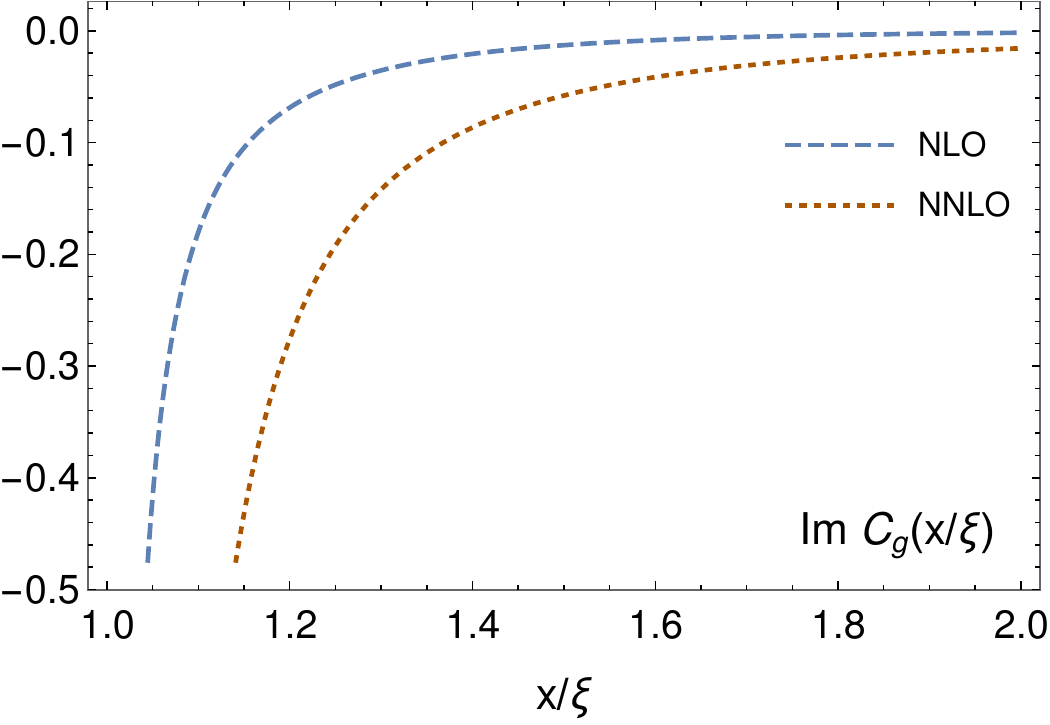}
\caption{The real (left panels) and imaginary (right panels) parts of the 
CFs $C_{u}$ and $C_g$ (\ref{eq:CFF-H}) as functions of $x/\xi$ 
at $\mu^2 = Q^2 = 4\text{ GeV}^2$ for $n_f=3$ and $\alpha_s(4\text{ GeV}^2) = 0.29751$ for NLO, 
$\alpha_s(4\text{ GeV}^2) = 0.300224$ for NNLO.
Solid lines: LO (black), short dashes: NLO (blue), long dashes: NNLO (orange).
The delta-function contributions to the imaginary parts at $x=\xi$ 
are  not shown. 
}
\label{figCF}
\end{figure*}

\end{widetext}

\end{document}